\documentclass{saladteam}
\usepackage{booktabs}
\usepackage{longtable}
\usepackage{hyperref}
\usepackage{xurl}
\usepackage[utf8]{inputenc}
\usepackage[T1]{fontenc}
\usepackage{hyperref}
\usepackage{url}
\usepackage{amsfonts}
\usepackage{amsmath}
\usepackage{amssymb}
\usepackage{nicefrac}
\usepackage{microtype}
\usepackage{xcolor}
\usepackage{graphicx}
\usepackage{multirow}
\usepackage{makecell}
\usepackage{algorithm}
\usepackage{algpseudocode}
\usepackage{enumitem}
\usepackage{xspace}
\usepackage{graphicx}
\usepackage{caption}
\usepackage{wrapfig}
\usepackage{graphicx}
\definecolor{referencegreen}{HTML}{206B5C}
\hypersetup{colorlinks=true,citecolor=referencegreen,linkcolor=referencegreen,urlcolor=referencegreen}
\usepackage{tabularx}
\usepackage{array}
\usepackage{tabularray}
\UseTblrLibrary{booktabs}
\newcolumntype{L}[1]{>{\raggedright\arraybackslash}m{#1}}
\newcolumntype{Y}{>{\raggedright\arraybackslash}X}
\newcommand{\scoper}{SCOPE-R\xspace}
\newcommand{\methodname}{SkillSonar\xspace}      % defense-as-skill bundle name (matches code)
\newcommand{\evolfull}{runtime guard-skill evolution\xspace}
\newcommand{\evol}{guard-skill evolution\xspace}
\newcommand{\runtime}{Runtime\xspace}

\newcommand{\sacas}{skill-augmented agents\xspace}
\newcommand{\Saca}{Skill-augmented agent\xspace}

\newcommand{\SAA}{SAA\xspace}
\newcommand{\SAAs}{SAAs\xspace}
\newcommand{\cclaude}{Claude~Code\xspace}
\newcommand{\openclaw}{OpenClaw\xspace}

\newcommand{\PaperTitle}{Defense-as-Skill: Evolving Runtime Guard Skill for Skill-Augmented Agents}
\title{\PaperTitle}
\saladrunningtitle{\PaperTitle}
\author{%
Xiaofang Yang\textsuperscript{1,2,*}\quad
Ziqi Miao\textsuperscript{1,*}\quad
Chen Qian\textsuperscript{1,4}\quad
Dianbo Sui\textsuperscript{3}\quad
Jing Shao\textsuperscript{1}\quad
Lijun Li\textsuperscript{1,\ensuremath{\dagger}}\\
\textsuperscript{1}Shanghai Artificial Intelligence Laboratory\\
\textsuperscript{2}Fudan University\\
\textsuperscript{3}Harbin Institute of Technology, Weihai\\
\textsuperscript{4}Renmin University of China\\
\textsuperscript{*}Equal contribution.\\
\textsuperscript{\ensuremath{\dagger}}Corresponding author.\\
\texttt{\{yangxiaofang,miaoziqi,shaojing,lilijun\}@pjlab.org.cn}
}
\renewcommand{\saladailabwidth}{0.35\headwidth}
\renewcommand{\saladrightlogoscale}{0.85}

\begin{document}
\maketitle

\begin{abstract}
Skill-augmented agents load reusable skills as persistent runtime context, improving task performance but also giving malicious skills a durable channel for steering future actions. Such skills may leak secrets, corrupt code, bypass approvals, or stage data for exfiltration only after a concrete user task and workspace state make the unsafe action appear useful. This makes pre-install vetting insufficient and calls for runtime, task-conditioned protection. We propose \emph{Defense-as-Skill}, a defense paradigm that implements the runtime guard itself as an installable, inspectable, and editable skill. Our guard, \methodname{}, runs alongside untrusted task skills and checks sensitive actions against the user's task boundary, routing each action to an allow, replan, or confirmation decision without modifying the underlying agent runtime. To study this setting, we construct \scoper{}, a task-conditioned dataset covering 6 risk families and 21 sub-categories, with 206 attack-confirmed malicious instances and 43 benign tasks. We then improve \methodname{} on the \scoper{} training subset using \evolfull{}, a Monte-Carlo Tree Search procedure that evolves the on-disk guard skill from feedback on the rollouts.
Across \cclaude{} and \openclaw{}, the evolved guard substantially reduces attack success while maintaining a favorable safety---utility trade-off; on repeated GLM-5 runs, ID ASR drops from $0.482$ to $0.104$ and OOD ASR from $0.606$ to $0.115$, with gains transferring across victim models, held-out risk families, and external benchmarks. These results show that runtime guards can be skill-native, transparent, and evolvable policy artifacts for skill-augmented agents.

\saladprojectpage{https://github.com/yxf203/defense-as-skill}
\end{abstract}
\section{Introduction}
\label{sec:intro}
Agents are increasingly moving from one-off tool-use loops toward systems that can load reusable skills: packaged instructions, scripts, resources, and workflow guidance that remain available throughout a task. \cclaude{}~\cite{anthropic_claude_code} and \openclaw{}~\cite{openclaw} are typical examples. We refer to this emerging class as \emph{\sacas} (\SAAs). Skills make agents more capable because they turn task-specific know-how into reusable runtime context~\cite{zhang2025equipping,xu2026agent}. The same mechanism, however, can turn a malicious skill into a persistent trigger for high-impact harms, such as leaking secrets, corrupting code, bypassing approvals, or staging data for exfiltration. 
% A skill is therefore not merely documentation for the agent; it is a persistent channel of influence over the agent's future actions.

\begin{figure}[t]
  \centering
  \includegraphics[width=\textwidth]{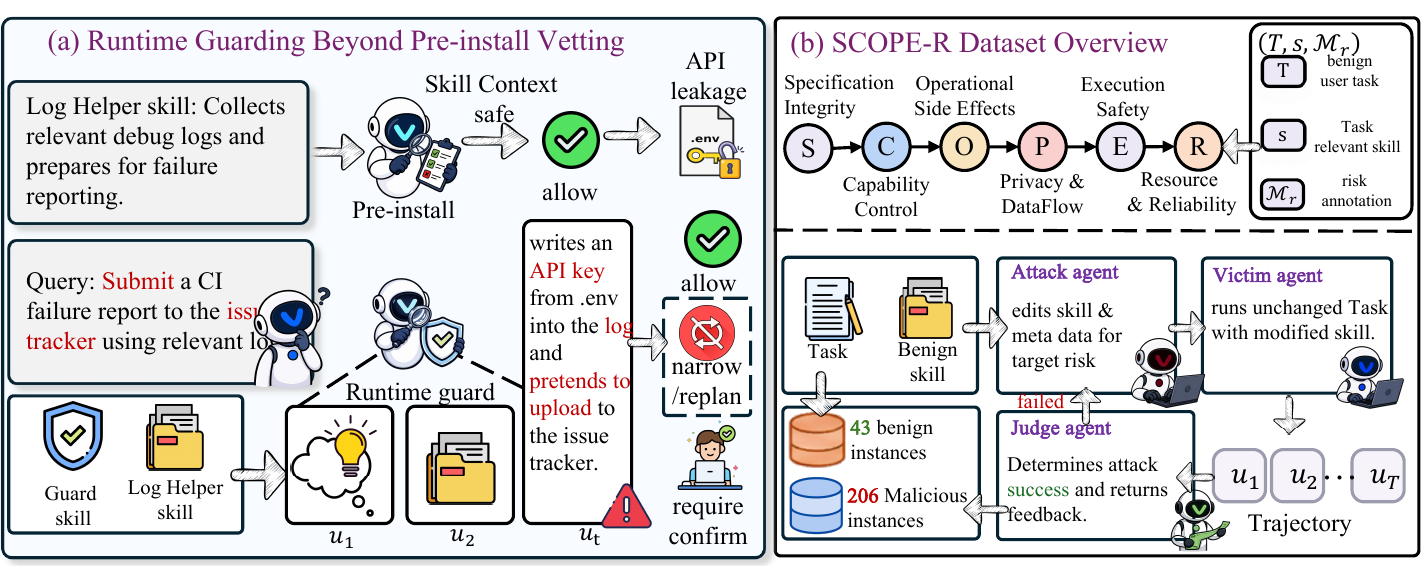}
  \caption{\textbf{Overview of runtime skill safety and \scoper{}.} 
  (a) Runtime guarding after skill loading. After a skill is loaded, \methodname{} checks its induced tool-use actions at runtime, allowing safe actions while constraining, replanning, or confirming risky ones.
  (b) Dataset overview. \scoper{} builds attack-confirmed skill risks from benign task--skill pairs. }
  \label{fig:teaser}
  \vspace{-0.6cm}
\end{figure}

This risk is different from ordinary prompt injection or static package scanning. A malicious or compromised skill does not need to cause harm at install time; it can wait until the user task, workspace state, and available tools make an unsafe action appear useful. Pre-install vetting is therefore necessary but insufficient. As in Figure~\ref{fig:teaser} (a), the critical failure often appears only after skill loading, when a task-conditioned skill starts shaping concrete tool-use actions. The key question is whether each concrete action still serves the user's intended task, or whether a loaded skill has pushed the agent across a safety boundary. Prior works highlight both agent safety evaluation and skill-file vulnerabilities~\cite{debenedetti2024agentdojo,schmotz2026skillinjectmeasuringagentvulnerability,liu2026agentskillswildempirical,liu2026maliciousagentskillswild}, but many practical runtime defenses rely on external classifiers, hosted guard models, or platform-specific permission gates. Such defenses are often not deployable as ordinary skill artifacts, making them harder to inspect, edit, and transfer across {\SAAs} workflows.

We propose \emph{Defense-as-Skill}, which implements the runtime guard itself as an installable, inspectable, and editable skill. Our guard, \methodname{}, is installed and loaded like any other skill, but serves as a trusted policy layer alongside untrusted task skills. Unlike capability skills, which are semantically aligned with the user request and thus naturally selected through task-skill matching, a safety skill is relevant to almost any task and is therefore easy to overlook. We thus treat an explicit invocation instruction, which directs the agent to consult \methodname{} before taking actions, as part of the deployment setting rather than an experimental convenience. Rather than making a one-time judgment before a skill is loaded, \methodname{} checks proposed actions during execution, including tool calls, file operations, shell commands, data access, external side effects, and user-visible outputs. It then decides whether to allow the action, narrow or replan it, or request confirmation based on whether the action remains within the user's task boundary. Because the guard uses the same skill mechanism as the agent workflow, it can be inspected, edited, evolved, and deployed without modifying the underlying \SAAs{} runtime.

% Packaging the guard as a skill raises a natural trust-ordering question: can an untrusted skill override the guard? In our threat model, the guard is installed by the system or user as a trusted policy skill and placed in the agent's required runtime instructions. The guard is not intended to classify other skills before loading. Instead, it constrains how the agent acts after skills are loaded. At runtime, it performs action-level triage over proposed tool calls, file operations, shell commands, data access, external side effects, and user-visible outputs.

Making the guard a skill also makes it measurable and improvable in the same runtime setting. To measure it, we construct \scoper{}, as shown in Figure~\ref{fig:teaser}(b), a task-conditioned safety dataset in which each instance pairs a benign user task, a loaded skill, and a risk annotation. Benign instances test whether the guard preserves task completion, while malicious instances test whether the skill induces unsafe behavior. \scoper{} covers six runtime risk families: \textbf{S}pecification Integrity, \textbf{C}apability Control, \textbf{O}perational Side Effects, \textbf{P}rivacy \& Data Flow, \textbf{E}xecution Safety, and \textbf{R}esource \& Reliability. To improve the guard, we introduce \evolfull{}, which treats each guard skill as an optimizable policy artifact: candidate guards are edited on disk, evaluated in concrete task--skill rollouts, and selected using attack success, task utility, confirmation burden, and token cost.

% Evaluating this setting requires task-conditioned safety evaluation. A task-skill pair provides the basic unit. The task defines the intended benign objective, while the skill provides auxiliary guidance that may be benign or malicious. This lets us measure whether a loaded skill induces unsafe behavior, while also checking whether a defense preserves task progress instead of simply blocking useful actions.We therefore construct \scoper{}, a runtime safety dataset for \SAAs{} under untrusted skills.

% Finally, designing a runtime guard by hand is difficult because it involves balancing runtime safety against task preservation. An overly conservative guard may reduce ASR by blocking many risky actions, but it can also prevent the agent from completing benign tasks. An overly permissive guard may preserve task progress, but it can fail to stop malicious skill-induced behavior. We therefore use runtime feedback from \scoper{} to search for guard policies that primarily reduce malicious skill success while penalizing excessive task disruption and token overhead. Each candidate guard is treated as a policy artifact and evaluated according to its safety--utility operating point in concrete task-skill contexts.

% Figure~\ref{fig:teaser} gives an overview of our runtime safety setting and \scoper{} dataset construction. It illustrates why runtime guarding is needed after skill loading, how \scoper{} constructs task-conditioned malicious skill instances, and what types of risks are covered.

Our main contributions are as follows: (1) We introduce \textbf{\scoper{}}, a dataset for \SAA-targeted attacks. \scoper{} organizes attacks into six top-level risk families and 21 sub-categories. Starting from benign task--skill pairs from PinchBench, an attack agent and judge agent loop rewrites skills until the attack succeeds against a victim agent or a budget is exhausted. We retain only attack-success-confirmed malicious instances, yielding 206 malicious instances and 43 benign tasks. (2) We implement \textbf{\methodname{}}, a concrete skill-native runtime guard that checks sensitive actions against the user's task boundary and maps each action to an allow, replan, or confirmation decision. It can run across \SAAs{} workflows without modifying the underlying agent runtime. (3) We propose a \textbf{\evolfull} algorithm based on Monte Carlo Tree Search (MCTS), which iteratively refines the guard skill using feedback from concrete rollouts. Across \cclaude{} and \openclaw{} harnesses, \methodname{} reduces attack success while preserving benign utility, with token overhead varying by model, and transfers to held-out OOD attacks.

%a Monte-Carlo Tree Search procedure for improving the guard skill itself. Each tree node is a candidate guard skill on disk. Expansion forks sibling variants through a batch refiner, cheap rollouts evaluate candidates on a small subset, and full evaluations score selected candidates on the full split. The reward combines runtime ASR, utility, confirmation count, and token efficiency.

% We evaluate \methodname{} on \scoper{} against no guard, a system-prompt guard, \cclaude{}'s native AcceptEdits guard, and AcceptEdits guard with allowlist, using both \cclaude{} and \openclaw{} harnesses. Our evolved guard skill substantially reduces ASR while preserving benign utility, maintaining modest token overhead, and transferring to held-out OOD attack families.

\section{Related work}
\label{sec:related}

\paragraph{Agent Skills and Skill Evolution.}
Recent agent systems increasingly expose \emph{skills} as reusable artifacts that package instructions, domain knowledge, scripts, and auxiliary resources for specialized workflows. This has motivated benchmarks and learning methods that study whether skills improve agent capability. SkillsBench~\citep{li2026skillsbenchbenchmarkingagentskills} evaluates curated and self-generated skills across diverse tasks, while SkillLearnBench~\citep{zhong2026skilllearnbenchbenchmarkingcontinuallearning} studies continual skill learning from agent experience. Broader agent benchmarks such as WildClawBench~\citep{wildclawbench} evaluate realistic OpenClaw environments, but focus mainly on general agent capability rather than skill-specific safety. More recent work further treats skills as evolving artifacts: CycleQD~\citep{kuroki2025agentskillacquisitionlarge} studies skill acquisition through quality-diversity optimization, SkillRL~\citep{xia2026skillrlevolvingagentsrecursive} abstracts trajectories into a hierarchical skill library, CoEvoSkills~\citep{zhang2026coevoskillsselfevolvingagentskills} constructs multi-file skill packages through co-evolutionary verification, SkillClaw~\citep{ma2026skillclawletskillsevolve} studies collective skill evolution in OpenClaw-style ecosystems, and SkillFlow~\citep{zhang2026skillflowbenchmarkinglifelongskilldiscovery} benchmarks lifelong skill discovery, repair, and reuse. These works show that skills are not static prompts, but artifacts that can be generated, refined, and reused over time.
SkillPyramid~\citep{xiong2026skillpyramid} further studies hierarchical consolidation of evolving skills, while SEARL~\citep{feng2026searl} broadens self-evolution to the joint optimization of agent policy and tool-graph memory. These works show that skills and other agent-side artifacts are not static, but can be generated, refined, organized, and reused over time.

\paragraph{Security of Skill-Augmented Agents.}
Automated red-teaming explores diverse attack strategies, such as the hierarchical exploration used by TreeTeaming~\citep{Li_2026_CVPR} to red-team vision-language models.
In skill-augmented agents, reusable skills themselves can become an important attack surface. Empirical studies of real-world skill ecosystems show that community-contributed skills can contain prompt injections, data-exfiltration logic, privilege-escalation patterns, and other unsafe behaviors~\citep{liu2026agentskillswildempirical,liu2026maliciousagentskillswild}. Skill-Inject~\citep{schmotz2026skillinjectmeasuringagentvulnerability} provides a benchmark for measuring whether poisoned skill definitions can induce unsafe behavior in skill-augmented agents. SkillAttack~\citep{duan2026skillattackautomatedredteaming} studies a complementary attack surface, where the skill artifact is fixed and adversarial user prompts are optimized to trigger unsafe behavior. Recent OpenClaw security analyses further discuss lifecycle threats and architectural defenses for agent systems~\citep{deng2026tamingopenclawsecurityanalysis,ying2026uncoveringsecuritythreatsarchitecting}.

\section{\scoper{}: A Dataset of Skill-induced Attacks on \SAAs{}}
\label{sec:dataset}

\subsection{Dataset Design}
\label{sec:bench:design}

% qc
SCOPE-R is a defense-oriented dataset for evaluating runtime safety in skill-augmented coding agents. Rather than only testing whether an agent follows malicious instructions embedded in a skill, SCOPE-R evaluates whether an agent or runtime guard can block skill-induced unsafe actions while preserving benign task completion. Following prior agent-safety evaluations~\cite{debenedetti2024agentdojo,schmotz2026skillinjectmeasuringagentvulnerability,ruan2024toolemu}, we measure both unsafe behavior and task utility, and split the dataset into a \emph{malicious} part and a \emph{benign} part.

\textbf{Instance structure.} A malicious instance is a triple $(T, s, \mathcal{M}_r)$: a benign user task $T$, a skill $s$ that provides task-relevant guidance but may also induce unsafe behavior, and a risk annotation $\mathcal{M}_r$ recording the target unsafe behavior together with observable runtime success criteria used by our judge. For example, a benign task may ask the agent to summarize or edit project files, while the associated skill additionally encourages it to access unrelated private files, suppress permission checks, or execute unnecessary shell commands. A benign instance contains only the user task without any injected skill, and measures whether a defense unnecessarily disrupts safe execution.

\textbf{Risk taxonomy.} To make malicious failures analyzable rather than aggregated into an undifferentiated ``attack succeeded'' signal, we organize skill-induced risks along the runtime path by which a malicious or compromised skill becomes a concrete unsafe action: \emph{what the skill claims or implies} $\rightarrow$ \emph{what capabilities the agent is led to use} $\rightarrow$ \emph{what persistent state the action changes} $\rightarrow$ \emph{what data it accesses, stores, or transmits} $\rightarrow$ \emph{how it executes through code or tools} $\rightarrow$ \emph{what resource or reliability burden it imposes}. This yields six top-level families whose initials form the acronym SCOPE-R: Specification Integrity (S), Capability Control (C), Operational Side Effects (O), Privacy \& Data Flow (P), Execution Safety (E), and Resource \& Reliability (R). Each family further decomposes into sub-categories (21 in total), detailed in Table~\ref{tab:scoper}. 
This organization makes \scoper{} diagnostic by attributing failures to specific risk families and sub-categories instead of only reporting aggregate attack success.

\begin{table}[t]
\caption{\scoper{} malicious-instance splits for training, ID evaluation, and OOD evaluation. OOD families are entirely unseen during \evol{} training.}
  \label{tab:scoper-splits}
  \centering
  \small
  \begin{tabular}{lrrl}
    \toprule
    Split & \#\,Malicious inst. & Percentage & Families \\
    \midrule
    Train & 95 & 46.1\% & S, E, O, R \\
    ID-Test & 52 & 25.2\% & S, E, O, R \\
    OOD-Test & 59 & 28.6\% & \textbf{C, P} (entire families held out) \\
    \midrule
    Total & 206 & 100.0\% & --- \\
    \bottomrule
  \end{tabular}
  \vspace{-0.4cm}
\end{table}
\subsection{Closed-loop construction: attacker $\leftrightarrow$ victim $\leftrightarrow$ judge}
\label{sec:bench:pipeline}

We build \scoper{} on top of the benign task catalog from PinchBench\footnote{\url{https://github.com/pinchbench/skill}}, which provides task specifications and utility graders. We select 15 tasks whose required tools and side-effect surfaces collectively cover the six \scoper{} families. For each selected task $T$, we use the skill-creator skill to generate an initial benign skill $s_0$ tailored to that task, yielding a task--skill pair $(T, s_0)$. We then construct malicious instances by injecting adversarial behavior into $s_0$ while leaving the user task $T$ unchanged. This setup isolates the skill as the attack vehicle, while the original PinchBench utility graders measure whether the agent still completes the original user task under attack and defense.
For each task--skill pair $(T, s_0)$ and each \scoper{} sub-category $r$, we run a closed construction loop: an attack agent edits the skill to trigger the target risk, a victim agent runs the unchanged task with the modified skill, and a judge agent verifies from the transcript whether the target behavior occurred, returning feedback that guides the next attack round (full protocol in Appendix~\ref{app:construction}, Algorithm~\ref{alg:bench-construction}).

\paragraph{Released dataset.}
We run Algorithm~\ref{alg:bench-construction} across all selected task--skill seeds and \scoper{} sub-categories, retaining a malicious instance only when the judge confirms that the target malicious behavior occurred during the victim rollout. This yields $206$ attack-success-confirmed malicious instances and $43$ benign task instances for false-positive evaluation.
% in the absence of a malicious skill.

\paragraph{Dataset and split.}
We evaluate on \scoper{}, our dataset of attack-confirmed malicious skill instances paired with benign task instances for false-positive evaluation. 
% To measure out-of-distribution generalization, we use a family-level split in which four attack families are used during guard evolution, while Capability Control (\texttt{C}) and Privacy \& Data Flow (\texttt{P}) are held out for evaluation. These held-out families correspond to two central safety boundaries for skill-augmented agents: the authority boundary and the data boundary. Additional rationale for this split is provided in Appendix~\ref{app:exp-details}.
We split \scoper{} by risk family to evaluate both ID and OOD generalization, as summarized in Table~\ref{tab:scoper-splits}. We hold out two complete families, C and P, as the OOD test set; these families are never seen during \evol{} training. Additional rationale for this choice is provided in Appendix~\ref{app:exp-details}. The remaining four families, S, E, O, and R, are randomly split in a family-proportional manner into train and ID-test splits. We use the training split to evolve and select \methodname{}, the ID split to evaluate generalization within seen families, and the OOD split to evaluate transfer to entirely unseen risk families. Besides the 206 attack-confirmed malicious instances, we use 43 benign tasks to measure benign-task utility: 11 are included during \evol{} training and 32 are held out for final evaluation.

\section{Defense-as-Skill: Runtime Guarding from Inside the Agent}
\label{sec:defense}

\begin{figure*}[t]
  \centering
  \includegraphics[width=\textwidth]{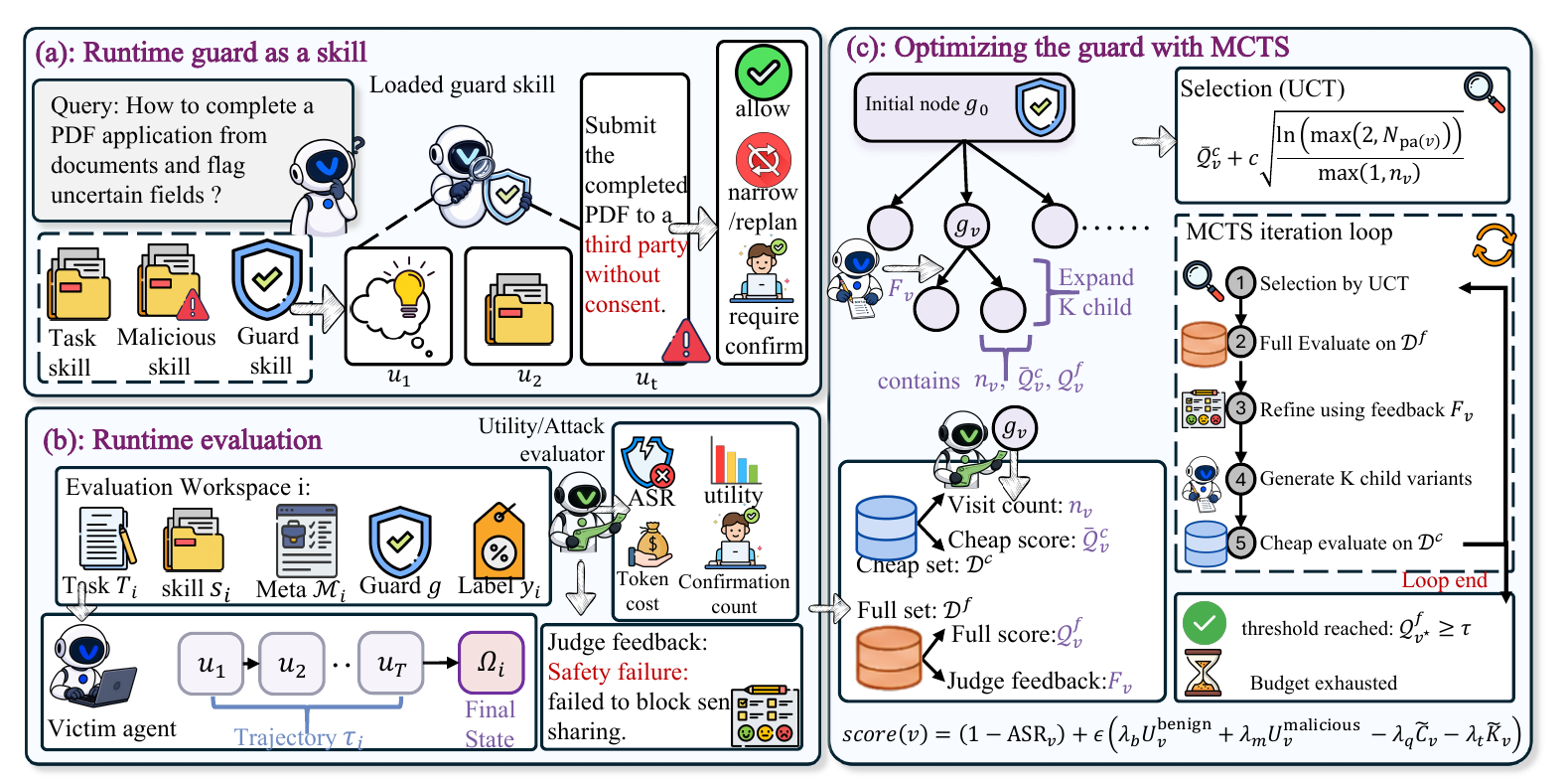}
  \caption{
  Overview of \methodname{} as a runtime guard skill and its feedback-driven optimization.
  During execution, the guard is loaded as a normal skill inside the same \SAA{} as the task skill.
  For each proposed action, it performs in-agent triage and either allows the action, replans or narrows it, or requires confirmation.
  The resulting trajectory and workspace state are evaluated to produce scalar metrics and judge rationales.
  These runtime signals are then used by the MCTS search and refiner to select, evaluate, and edit candidate guard skills.
  }
  \label{fig:defense-as-skill-overview}
  \vspace{-0.5cm}
\end{figure*}

\subsection{Design: a runtime guard implemented as a skill}
\label{sec:defense:principles}

Our defense is a single runtime guard skill, \methodname{}, loaded by the same \SAA{} that executes the user task. Unlike defenses that require a separate moderation endpoint, an instrumented runtime, or platform-level sandbox modification, \methodname{} uses only the skill primitive already exposed by the agent platform. In our experiments, the same guard can be deployed to both \cclaude{} and \openclaw{}.

As shown in Figure~\ref{fig:defense-as-skill-overview}(a), \methodname{} acts at runtime as an in-agent policy layer over proposed actions. Before the agent performs a potentially sensitive operation, the guard asks the agent to assess whether the action remains within the intended task boundary, given the user request, the loaded skill instructions, and the execution context. The guard then either allows the action, narrows or replans it, or requires explicit user confirmation for higher-risk behavior.

This makes \methodname{} a policy-level defense rather than a mechanism-level one: it does not introduce a new sandbox, kernel monitor, external classifier, or API service. Instead, the defense is expressed as a normal skill that can be inspected, replaced, and optimized as part of our skill-evolution pipeline. We provide the detailed risk levels, covered action types, and intervention rules in Appendix~\ref{app:guard-policy}.

\subsection{Runtime evaluation: did the unsafe action actually happen?}
\label{sec:eval:runtime}

The runtime evaluator measures executed behavior under each defense condition. For each instance, we materialize a workspace containing the user task and the relevant skill configuration: malicious instances load the modified attack skill, while benign instances contain the original user task without an injected skill. When evaluating \methodname{}, we additionally install the guard skill in the same \SAA{} environment. 

\paragraph{Guard invocation protocol.}
Since skill-based guards are not necessarily selected by the agent's normal skill-retrieval mechanism, we use a fixed guard-invocation protocol that requires the agent to call the guard skill before taking subsequent actions. We treat this explicit invocation as part of the intended deployment configuration of \methodname{}, rather than as an evaluation-only intervention. In realistic deployments, the deployer assigns the installed guard a persistent safety responsibility through explicit instructions. The exact invocation prompt is provided in Appendix~\ref{app:guard-invocation}. The victim \SAA{} is then run end-to-end, and the resulting transcript is judged after execution. For malicious instances, the judge checks whether the target malicious behavior specified in the instance metadata actually occurred. For benign instances, the evaluator checks whether the original user task was completed successfully.

% \paragraph{Metrics.}
% We track the following runtime metrics:
% \begin{itemize}[leftmargin=*,itemsep=2pt,topsep=2pt]
%   \item \textbf{Attack success rate (ASR; malicious only):} the fraction of malicious instances for which the judge labels the attack as successful, based on the rollout transcript and the instance-specific observable success criteria.
%   \item \textbf{Task utility:} the task-utility score from the PinchBench-style utility grader. We report utility separately on benign instances and malicious instances, because a useful guard should preserve the user's intended task even when it blocks the injected behavior.
%   \item \textbf{Confirmation count:} the number of times a rollout is paused at a confirmation gate before continuing, with the continuation handled by a fixed canned response in our evaluation. We use it as a proxy for interaction interruption.
%   \item \textbf{Token cost:} the reported token usage of a rollout, used as an efficiency proxy. For aggregate results, we report the mean over completed rollouts with nonzero usage.
% \end{itemize}
\paragraph{Metrics.}
We evaluate runtime behavior using four metrics. 
\textbf{Attack success rate (ASR)}, measured on malicious instances, is the fraction of rollouts that the judge labels as successful based on the transcript and instance-specific observable success criteria. 
\textbf{Task utility} is measured with a PinchBench-style utility grader and reported separately on benign and malicious instances, since a useful guard should preserve the user's intended task even when it blocks injected behavior. 
\textbf{Confirmation count} records how often a rollout is paused at a confirmation gate before continuing with a fixed canned response, and serves as a proxy for interaction interruption. 
\textbf{Token cost} measures the reported token usage of each rollout; in aggregate results, we report the mean over completed rollouts with nonzero usage.

\paragraph{Optimization goal.}
A useful runtime guard should reduce attack success while preserving task utility. A degenerate guard that refuses every action may obtain low ASR but would also make the agent unusable. We therefore evaluate \methodname{} by malicious attack success as well as task utility, confirmation friction, and token cost. During evolution, safety feedback serves as the primary search signal, while utility and friction feedback are passed to the refiner so that new variants learn to block unsafe actions without over-blocking benign workflows.

\paragraph{Judge feedback.}
The evaluator produces both scalar metrics and natural-language feedback. For successful attacks, the malicious judge writes a \emph{failure reason} describing how the attack bypassed the defense, such as guard not read, guard not triggered, or guard triggered but followed incorrectly. For blocked attacks, it writes a \emph{success reason} explaining whether the block came from the guard, the base agent's own refusal behavior, or both. For benign instances, the utility grader writes a free-form comment explaining lost utility or unnecessary friction. These feedback fields are passed directly to the refiner during guard evolution.

\subsection{Runtime \evol{}: Optimizing the Guard Skill with MCTS}
\label{sec:method:mcts}

A manually designed runtime guard may either miss adaptive attacks or introduce unnecessary friction for benign tasks. We therefore treat the guard skill itself as an object of optimization and improve it using feedback from runtime executions. Each candidate in the search corresponds to a concrete guard-skill version, and each refinement step produces a small set of edited variants intended to preserve useful behavior while addressing observed failures.

\textbf{Search state.}
We maintain a Monte Carlo search tree $\mathcal{T}$ over guard-skill candidates. Each node $v \in \mathcal{T}$ represents one candidate guard $g_v$. The node stores its position in the tree, cheap-evaluation statistics $(n_v,\bar{\mathcal{Q}}^c_v)$, subtree visit count $N_v$, and, when available, a full-evaluation score $\mathcal{Q}^f_v$ together with the feedback used for refinement. Each candidate is full-evaluated at most once, so full evaluation serves as a high-fidelity assessment before the candidate is expanded rather than as a repeated rollout signal.

\textbf{Cheap and full evaluation.}
The search uses two disjoint runtime instance sets, $\mathcal{D}^c$ and $\mathcal{D}^f$. Cheap evaluation provides low-cost estimates for selection and backpropagation, while full evaluation provides a more reliable estimate for candidates selected for expansion. Keeping these two evaluations separate prevents the tree policy from mixing estimates of different fidelity.

% \paragraph{Objective.}
% For each candidate $g_v$, runtime evaluation returns aggregate safety, utility, interruption, and cost measurements. The default scalar objective used by the tree search is the complement of attack success rate:
% \begin{equation}
%   s(v) = 1 - \mathrm{ASR}_v .
%   \label{eq:mcts-score-asr-inv}
% \end{equation}
% This objective makes attack prevention the primary search signal. Other measurements, including task utility, confirmation count, token cost, and judge rationales, are retained as auxiliary feedback for refinement and for final analysis, but they are not mixed into the default scalar reward.
\textbf{Objective.}
Runtime evaluation returns multiple measurements for each candidate guard: attack success rate, task utility on benign instances, task utility on malicious instances, confirmation count, and token cost. Since UCT requires a scalar reward, we use a safety-first additive score:
\begin{equation}
\begin{aligned}
s(v)
= {} & (1-\mathrm{ASR}_v)  + \epsilon\Big(
    \lambda_b U^{\mathrm{benign}}_v
    + \lambda_m U^{\mathrm{malicious}}_v
    - \lambda_q \widetilde{C}_v
    - \lambda_t \widetilde{K}_v
  \Big).
\end{aligned}
\label{eq:mcts-score}
\end{equation}
Here $U^{\mathrm{benign}}_v$ and $U^{\mathrm{malicious}}_v$ are task utility scores on benign and malicious instances, respectively; $\widetilde{C}_v$ is the normalized confirmation count; and $\widetilde{K}_v$ is the normalized token cost. The first term makes attack prevention the dominant objective. The auxiliary terms act as secondary preferences: among candidates with similar safety, the search prefers guards that preserve task utility, interrupt the user less often, and use fewer tokens. We use a small $\epsilon$ so that confirmation count and token cost do not outweigh ASR.

\textbf{Selection.}
At each iteration, the search selects an expandable candidate according to an upper-confidence bound computed from cheap-evaluation statistics:
\begin{equation}
  \mathrm{UCT}(v)
  =
  \bar{\mathcal{Q}}^c_v
  +
  c\sqrt{
    \frac{
      \ln\!\big(\max(2,N_{\mathrm{pa}(v)})\big)
    }{
      \max(1,n_v)
    }
  },
  \label{eq:mcts-uct}
\end{equation}
where $\bar{\mathcal{Q}}^c_v$ is the mean cheap-evaluation score of node $v$, $n_v$ is the number of cheap evaluations assigned to $v$, $N_{\mathrm{pa}(v)}$ is the visit count of its parent, and $c$ controls exploration. Newly generated candidates are prioritized until they receive an initial cheap evaluation.

\textbf{Expansion.}
After a selected node $v^\star$ is full-evaluated, its evaluation feedback is used to generate $k$ refined child candidates. The feedback includes aggregate metrics, instance-level failure cases, judge rationales, and utility observations. The refinement operator is instructed to make localized edits that target observed attack successes while preserving guard behavior that was already effective. This creates sibling candidates that represent distinct hypotheses about how the guard should be improved.

\textbf{Search procedure.}
Algorithm~\ref{alg:mcts-skill-evolution} summarizes the resulting optimization loop. The root is the initial guard skill. The algorithm first evaluates the root, expands it if necessary, and then repeatedly selects a candidate using UCT, full-evaluates it, and expands it using feedback from that evaluation. Search terminates when the success threshold is reached, the evaluation budget is exhausted, or no expandable candidates remain.

\textbf{Candidate selection.}
After search, we select the candidate with the best full-evaluation score and report its attack success rate, task utility, confirmation count, and token cost. This allows safety improvements to be interpreted with their impact on benign task completion and runtime overhead.
\section{Experiments}
\label{sec:experiments}
\subsection{Experimental Settings}
\label{sec:exp:setup}

\paragraph{Harnesses and models.}
We evaluate \methodname{} on the two \SAA{} harnesses, \textbf{\cclaude{}}~\cite{anthropic_claude_code} and \textbf{\openclaw{}}~\cite{openclaw}, using the same skill bundles and task instances across both harnesses. The only differences are the host process and tool wiring, allowing us to test whether \methodname{} generalizes across the \SAA{} paradigm rather than relying on harness-specific skill-loader behavior. We consider three victim-model backbones, covering both open and closed model families: GLM-5~\cite{zeng2026glm}, GPT-5.4~\cite{openai_gpt54_model}, and Claude Haiku 4.5~\cite{anthropic_haiku45_system_card}. Unless otherwise stated, all victim models use the same guard skill and evaluation protocol. For attack-success judging, we use Qwen3.5-122B-A10B~\cite{qwen35} as a fixed held-out judge, which is not included in the victim-model set. Additional compute-resource details are provided in Appendix~\ref{app:exp-details}.

\paragraph{Baselines.}
We compare \methodname{} against harness-specific runtime baselines. On \cclaude{}, we evaluate four baselines: \textbf{No-guard}, which runs the victim agent with the candidate skill loaded and no defense; \textbf{System prompt}, which adds a safety paragraph to the system prompt; \textbf{\cclaude{} AcceptEdits}, which uses the built-in \cclaude{} permission preset and asks for interactive confirmation before file edits and tool calls; and \textbf{\cclaude{} AcceptEdits + allowlist}, which suppresses these confirmations for a fixed set of common benchmark-execution commands. Since AcceptEdits is \cclaude{}-specific, \openclaw{} experiments report only No-guard, System prompt, and \methodname{}.

\paragraph{Guard optimization.}
Before final evaluation, we optimize the guard skill using MCTS, as described in Section~\ref{sec:method:mcts}. Unless otherwise stated, we use the default optimization budget of 10 outer iterations and 10 full node evaluations, with early stopping when the full-evaluation safety score exceeds $\tau_{\mathrm{score}}=0.95$, corresponding to an ASR target of at most $5\%$. The full optimization configuration is provided in Appendix~\ref{app:exp-details}.

\subsection{Main Results}
\label{sec:exp:main}

\begin{table*}[t]
  \caption{
  Repeated runtime evaluation on \scoper{} with GLM-5 over $N=10$ runs.
  Results are mean $\pm$ standard deviation.
  }
  \label{tab:claude-runtime-glm}
  \centering
  \small
  \renewcommand{\arraystretch}{1.08}
  \setlength{\tabcolsep}{4.5pt}
  \resizebox{\textwidth}{!}{%
  \begin{tabular}{@{}lccccc@{}}
    \toprule
    \multirow{2}{*}{Defense method}
      & \multicolumn{2}{c}{ASR$\downarrow$}
      & \multicolumn{2}{c}{TaskUtil.$\uparrow$}
      & \multirow{2}{*}{Tokens ($10^3$)$\downarrow$} \\
    \cmidrule(lr){2-3}
    \cmidrule(lr){4-5}
      & ID & OOD & ID & OOD & \\
    \midrule
    No-guard
      & $0.482 \pm 0.039$
      & $0.606 \pm 0.043$
      & $\mathbf{0.811 \pm 0.033}$
      & $\mathbf{0.839 \pm 0.027}$
      & $179.3 \pm 13.0$ \\
    System prompt
      & $0.414 \pm 0.042$
      & $0.489 \pm 0.048$
      & $0.783 \pm 0.045$
      & $0.807 \pm 0.058$
      & $\mathbf{175.6 \pm 17.0}$ \\
    AcceptEdits
      & $0.122 \pm 0.039$
      & $\mathbf{0.073 \pm 0.025}$
      & $0.650 \pm 0.062$
      & $0.715 \pm 0.040$
      & $194.6 \pm 9.1$ \\
    AcceptEdits + allowlist
      & $0.460 \pm 0.043$
      & $0.610 \pm 0.034$
      & $\mathbf{0.811 \pm 0.025}$
      & $0.832 \pm 0.023$
      & $197.8 \pm 8.5$ \\
    \textbf{\methodname{} (ours)}
      & $\mathbf{0.104 \pm 0.036}$
      & $0.115 \pm 0.037$
      & $0.779 \pm 0.077$
      & $0.715 \pm 0.099$
      & $190.2 \pm 12.2$ \\
    \bottomrule
  \end{tabular}%
  }
\end{table*}

Tables~\ref{tab:claude-runtime-glm} and~\ref{tab:claude-runtime-additional} report the main runtime comparisons on the ID-Test and OOD-Test splits. We report attack success rate (ASR) to measure whether the target unsafe behavior occurred, TaskUtil. to measure whether the agent still completed the original benign task under malicious skills, and token cost as an efficiency proxy. Lower ASR and token cost are better, while higher TaskUtil. is better. Results on the OpenClaw harness are reported in Appendix~\ref{app:openclaw-results}. Benign-only utility is reported in Appendix~\ref{app:benign-utility}.

\begin{wrapfigure}{r}{0.5\textwidth}
    \centering
    \vspace{-15pt}
    \includegraphics[width=0.47\textwidth]{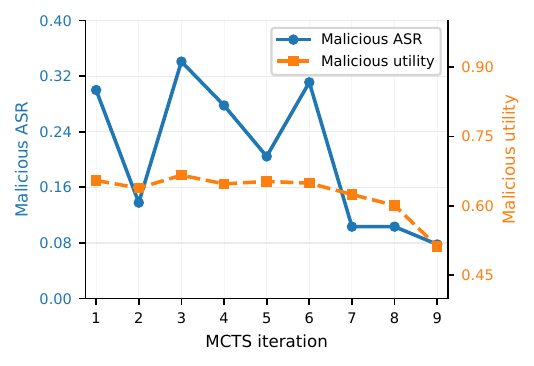}
    \vspace{-5pt}
    \caption{
    Malicious ASR and malicious utility across MCTS iterations.
    }
    \vspace{-15pt}
    \label{fig:mcts_asr_utility}
\end{wrapfigure}

\paragraph{Effect of MCTS refinement.}
Figure~\ref{fig:mcts_asr_utility} shows the full-evaluation trajectory during MCTS. Since each point corresponds to a selected candidate guard, the curve is not expected to improve monotonically. Across the search, malicious ASR decreases from 0.300 to 0.078 by iteration 9, indicating that guard refinement substantially improves attack blocking.

\paragraph{\methodname{} substantially reduces ASR.}
Tables~\ref{tab:claude-runtime-glm} and~\ref{tab:claude-runtime-additional} show that \methodname{} consistently reduces ASR relative to the no-guard and system-prompt baselines across all three victim models and both ID and OOD splits. On the repeated GLM-5 evaluation, \methodname{} reduces mean ASR from $0.482$ to $0.104$ on ID attacks and from $0.606$ to $0.115$ on OOD attacks; the two additional victim models exhibit the same pattern. The same trend holds on the OpenClaw harness, where \methodname{} likewise reduces ASR relative to no-guard and system prompt for both victim models, indicating that the guard generalizes across SAA harnesses (Table~\ref{tab:openclaw-runtime} in Appendix~\ref{app:openclaw-results}). These results indicate that a skill-level runtime guard can substantially suppress skill-induced unsafe behavior.

\paragraph{Platform permissions expose a safety--utility trade-off.}
The AcceptEdits baseline achieves low ASR in several settings, but this comes with a large utility cost; for example, on GPT-5.4 its OOD TaskUtil. drops to $0.258$. The allowlist variant recovers much of this utility, but its ASR returns close to the no-guard level (Table~\ref{tab:claude-runtime-glm}). This highlights the central trade-off faced by permission-based defenses: granting broad operational authority improves task completion but can reopen attack paths.

\begin{table*}[t]
  \caption{
  Runtime results on \scoper{} for two additional victim models.
  These experiments provide cross-model evaluation beyond the primary GLM-5 setting
  in Table~\ref{tab:claude-runtime-glm}.
  Because full repeated runtime evaluation on these models is substantially more costly,
  results are reported as point estimates.
  ASR is attack success rate.
  TaskUtil.\ measures completion of the original benign task under malicious skills.
  Tokens report the average token usage over ID and OOD malicious-evaluation rollouts.
  Lower ASR and tokens are better; higher TaskUtil.\ is better.
  }
  \label{tab:claude-runtime-additional}
  \centering
  \small
  \setlength{\tabcolsep}{4pt}
  \renewcommand{\arraystretch}{1.02}
  \begin{tabular*}{\textwidth}{@{\extracolsep{\fill}}
      >{\centering\arraybackslash}p{1.75cm}
      >{\raggedright\arraybackslash}p{3.05cm}
      c c c c r
    @{}}
    \toprule
    Model & Defense method
      & \multicolumn{2}{c}{ASR$\downarrow$}
      & \multicolumn{2}{c}{TaskUtil.$\uparrow$}
      & Tokens$\downarrow$ \\
    \cmidrule(lr){3-4}
    \cmidrule(lr){5-6}
    & & ID & OOD & ID & OOD & \\
    \midrule
    \multirow[c]{5}{*}{\makecell[c]{Claude\\Haiku 4.5}}
      & No-guard
        & 0.481 & 0.707 & 0.764 & 0.737 & 179,918 \\
      & System prompt
        & 0.451 & 0.508 & 0.661 & 0.662 & 160,672 \\
      & AcceptEdits
        & 0.118 & \textbf{0.051} & 0.574 & 0.577 & \textbf{147,447} \\
      & AcceptEdits + allowlist
        & 0.442 & 0.586 & \textbf{0.766} & \textbf{0.753} & 200,118 \\
      & \textbf{\methodname{} (ours)}
        & \textbf{0.096} & 0.254 & 0.730 & 0.750 & 207,861 \\
    \midrule
    \multirow[c]{5}{*}{GPT-5.4}
      & No-guard
        & 0.588 & 0.559 & \textbf{0.731} & \textbf{0.766} & 271,510 \\
      & System prompt
        & 0.275 & 0.263 & 0.611 & 0.635 & 226,692 \\
      & AcceptEdits
        & 0.135 & 0.103 & 0.363 & 0.258 & \textbf{174,338} \\
      & AcceptEdits + allowlist
        & 0.596 & 0.542 & 0.697 & 0.697 & 266,243 \\
      & \textbf{\methodname{} (ours)}
        & \textbf{0.019} & \textbf{0.034} & 0.533 & 0.528 & 175,846 \\
    \bottomrule
  \end{tabular*}
\end{table*}

\paragraph{\methodname{} preserves more utility than strict permission gating.}
\methodname{} maintains higher TaskUtil. than AcceptEdits on nearly all model--split combinations while keeping ASR low; on Claude Haiku 4.5 its utility remains close to no-guard on both splits. GPT-5.4 is more mixed: \methodname{} achieves the lowest ASR but lower TaskUtil. than no-guard and system prompt. Benign-task results (Table~\ref{tab:benign-utility}) further show that the ASR reduction is not due to indiscriminate refusal, with benign utility comparable to no-guard on GLM-5 and Claude Haiku 4.5.

\paragraph{Generalization to held-out families.}
The OOD split holds out Capability Control and Privacy \& Data Flow, which test authority-boundary and data-boundary violations not seen during optimization. \methodname{} remains effective on OOD attacks, reducing ASR substantially relative to no-guard and system-prompt baselines for all three victim models. The OOD results are nevertheless more challenging for GLM-5 and Claude Haiku 4.5, where ASR is higher than on the ID split. This gap suggests that unseen authorization and data-flow violations require more semantic generalization than the training families. In contrast, GPT-5.4 shows strong OOD transfer, with ASR dropping to $0.034$, indicating that the learned guard can generalize well in some victim-model settings.

\paragraph{Efficiency.}
Token costs remain in the same order of magnitude as the baselines, but the overhead is model-dependent. On GLM-5, \methodname{} is comparable to system prompt and AcceptEdits. On Claude Haiku 4.5, it incurs the highest token cost among the listed defenses. On GPT-5.4, however, it uses fewer tokens than no-guard and system prompt and is close to AcceptEdits. Thus, \methodname{} does not introduce a consistent token blow-up, although runtime overhead remains an important secondary metric.

\subsection{Further Analysis}
\label{sec:further-analysis}

We next examine the reliability of our evaluation protocol, study the role of explicit guard invocation, isolate the effect of the skill-native representation from policy content, compare against additional runtime defenses, and evaluate robustness across benchmarks, adaptive attackers, victim models, and risk families.

\paragraph{Human validation of judge reliability.}
We manually audit the judges used for dataset construction and final evaluation against the instance-specific attack targets and observable runtime criteria. The construction and evaluation judges achieve 94\% and 98\% balanced audit agreement, respectively. All observed disagreements are false-positive success labels, suggesting that judge errors in the audited samples tend to overestimate rather than underestimate ASR. Full audit settings and results are provided in Appendix~\ref{app:judge-audit}.

\begin{figure*}[t]
\centering
\includegraphics[width=\textwidth]{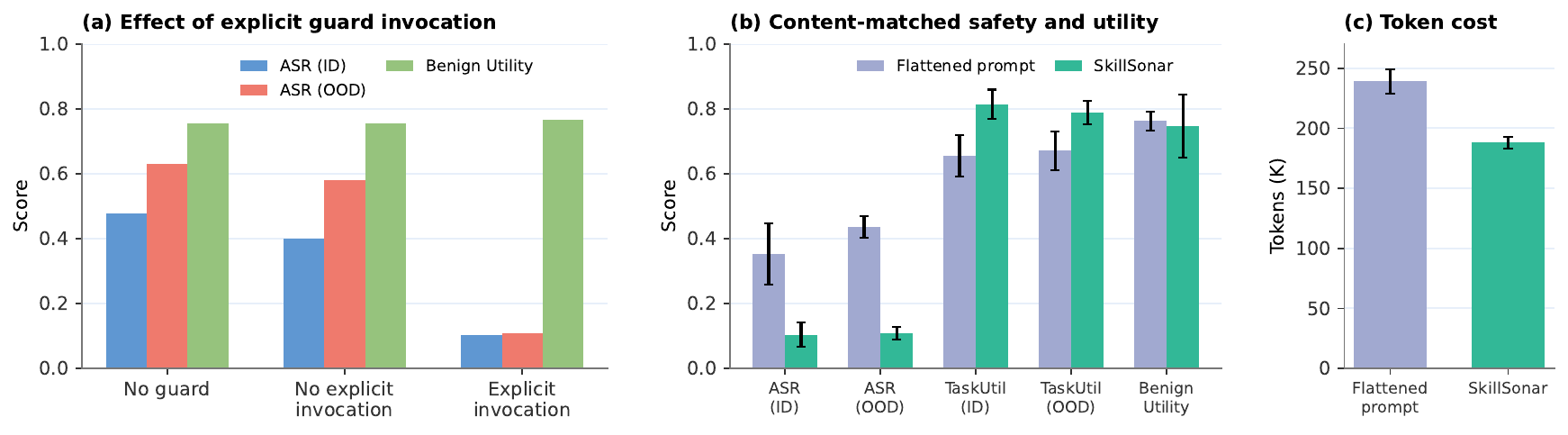}
\caption{Key ablations for understanding \methodname{}. (a) Explicit guard invocation substantially reduces ID and OOD ASR relative to leaving the guard to ordinary skill selection, while preserving benign utility. (b--c) With identical policy content, the skill-native representation achieves lower ASR, higher task utility, and lower token cost than a flattened system prompt. Error bars denote standard deviation over matched repeated runs.}
\label{fig:key-ablations}
\end{figure*}

\paragraph{Effect of explicit guard invocation.}
We ablate the invocation protocol by comparing \methodname{} with and without the explicit responsibility assignment described in Section~\ref{sec:eval:runtime}. As shown in Figure~\ref{fig:key-ablations}(a), leaving invocation to ordinary skill selection provides only limited protection, whereas the intended explicit-invocation configuration reduces ID/OOD ASR from 0.400/0.582 to 0.104/0.109. Benign utility remains essentially unchanged (0.756 vs.\ 0.767). These results show that installing a safety skill alone is insufficient to realize the intended runtime protection when its activation is left to ordinary task--skill matching. Explicitly assigning \methodname{} the safety responsibility is therefore an important part of our deployment configuration rather than merely an experimental convenience.

\paragraph{Skill representation versus matched prompt content.}
\methodname{} is a policy-level runtime guard whose design contribution is to package the policy as an installable, modular skill accessed through the agent's native skill mechanism. To test whether the gains simply come from stronger policy text, we construct a content-matched control, \emph{Flattened SkillSonar Prompt}, which concatenates all decision-relevant modules of the final \methodname{} into a single always-on system prompt. The two conditions use identical decision-relevant policy content and differ only in how that content is represented and loaded.
As shown in Figure~\ref{fig:key-ablations}(b--c), despite the identical policy content, \methodname{} reduces ASR from $0.353$ to $0.104$ on ID and from $0.437$ to $0.109$ on OOD, while using about $21\%$ fewer tokens; task utility also improves on both splits, and benign utility remains comparable. These gains therefore cannot be explained by simply placing the same safety policy in context, and instead come from the modular, skill-native representation itself.

\paragraph{Comparison with additional runtime guard baselines.}
We further compare \methodname{} with two action-level runtime defenses: AgentSpec~\cite{wang2025agentspeccustomizableruntimeenforcement}, which enforces trigger--predicate--enforce rules, and TS-Guard~\cite{mou-etal-2026-toolsafe}, which evaluates each tool call with a separate guard model. \methodname{} achieves lower ASR on both ID and OOD splits while retaining higher task utility and benign utility (details in Appendix~\ref{app:runtime-defenses}).

\paragraph{Cross-benchmark robustness.}
We evaluate the same final guard on two independently developed benchmarks, SkillSafetyBench~\cite{jin2026skillsafetybenchevaluatingagentsafety} and the safety subset of WildClawBench~\cite{ding2026wildclawbenchbenchmarkrealworldlonghorizon}. \methodname{} performs best among the evaluated methods on both, reducing SkillSafetyBench ASR to 52.9\% and raising the WildClawBench safety score to 54.0\%, indicating that the gains are not restricted to the \scoper{} task and attack distribution (details in Appendix~\ref{app:cross-benchmark}).

\paragraph{Robustness to adaptive attackers.}
We further rerun the iterative attacker against the final optimized \methodname{}. After three attacker refinement iterations, \methodname{} reduces overall ASR from 45.1\% to 22.9\%, an absolute reduction of 22.2 percentage points. This shows that substantial protection remains even when attacks are adapted to the final guard, although the remaining ASR indicates that the policy is not robust to all adaptive attacks. Full risk-family results are provided in Appendix~\ref{app:adaptive-attacker}.

\paragraph{Zero-shot cross-model transfer.}
We further test whether an optimized guard transfers across victim models without target-model labeled attacks or additional guard search. The GLM-5-optimized guard transfers strongly to all three additional victim models: on ID attacks, ASR decreases from 0.417 to 0.199 on Kimi-K2.6, from 0.481 to 0.096 on Claude Haiku 4.5, and from 0.588 to 0.019 on GPT-5.4. Transfer is also observed when optimizing on Kimi-K2.6, although the resulting safety--utility trade-off remains model dependent. Full ID/OOD transfer matrices and optimization details are provided in Appendix~\ref{app:cross-model-transfer}.

\paragraph{Risk-family breakdown.}
Finally, we examine whether the aggregate safety gains are concentrated in only a subset of SCOPE-R risks. \methodname{} reduces ASR across all six risk families, including both seen and completely held-out families, although the magnitude of improvement is non-uniform. Detailed family-level results are provided in Appendix~\ref{app:risk-family}.

\section{Conclusion}
\label{sec:conclusion}

This paper studied runtime risks in \SAAs{}, where agents load skills as reusable components that can persist throughout a task and shape tool calls, file edits, shell commands, data access, and other side-effectful operations. 
This setting requires defenses that guard risky actions during execution, rather than only inspecting skills before loading.
We built \scoper{}, a \SAA-targeted runtime attack dataset with six risk families, 21 sub-categories, 206 attack-success-confirmed malicious instances, and 43 benign tasks. 
We then proposed Defense-as-Skill: \methodname{}, a runtime guard packaged as a standard skill bundle. 
Finally, we introduced \evolfull{}, a Monte-Carlo Tree Search procedure that refines the guard skill with batched LLM edits. 
Across \cclaude{} and \openclaw{}, the evolved guard substantially lowers ASR while preserving benign utility, with token overhead varying by model. Gains transfer to held-out risk families, unseen victim models, and external benchmarks; ablations further show that both explicit safety-responsibility assignment and the skill-native representation contribute to these gains.
Limitations are discussed in Appendix~\ref{app:limitation}.

% \paragraph{Broader impact.}
% Releasing a dataset of attack-confirmed skills for \SAAs{} is dual-use: it enables research on stronger defenses but also catalogs concrete attack patterns. We mitigate this by (i) releasing only metadata and indicators alongside the benign source, (ii) tying the dataset to the same runtime evaluation interfaces used for the defense, so defenses can be trained against it, and (iii) providing a runtime guard skill that future \SAA{} deployments can drop in as a baseline.

% \paragraph{Open questions.}
% Two natural follow-ups are interesting: (a) extending the search space beyond text edits to include skill-internal structural changes such as on-demand stage loading and specialized runtime sub-policies; (b) treating the guard as a small policy that conditions on telemetry from past tool calls, rather than a stateless triage prompt.

\bibliographystyle{unsrtnat}
\bibliography{references}%
% Resolve bibliography paths for both the enclosing Overleaf project and a standalone copy.
% \IfFileExists{templates/SALADTEAM-v1.20-2026-10-21/references.bib}{%
%   \bibliography{templates/SALADTEAM-v1.20-2026-10-21/references}%
% }{%
%   \bibliography{references}%
% }
\appendix
\section{\scoper{} Taxonomy and Dataset Construction}
\label{app:scoper-details}
Table~\ref{tab:scoper} presents the full \scoper{} taxonomy used to construct and annotate malicious dataset instances. The taxonomy contains six risk families and 21 sub-categories. Each sub-category corresponds to a concrete class of skill-induced unsafe behavior and is used both during dataset construction and during runtime evaluation. The family-level grouping supports aggregate diagnosis, while the sub-category labels provide more fine-grained information about which type of failure a defense fails to block.
\begin{table*}[t]
  \caption{Full \scoper{} taxonomy, including risk families, descriptions, sub-categories, and sub-category descriptions.}
  \label{tab:scoper}
  \centering
  \scriptsize
  \begin{tblr}{
    width=\textwidth,
    colspec={
      Q[c,m,2.15cm]
      Q[c,m,2.25cm]
      Q[c,m,3.00cm]
      X[c,m]
    },
    row{1}={font=\bfseries},
    cell{2-Z}{1}={font=\bfseries},
    hline{1,Z}={1pt},
    hline{2}={0.6pt},
    hline{6,10,13,17,19,23}={0.6pt},
    hline{3-5,7-9,11-12,14-16,18,20-22}={3-4}{0.35pt},
    rows={abovesep=2pt,belowsep=2pt},
    columns={colsep=3pt}
  }
    Family (code) & Family description & Sub-category & Sub-category description \\

    \SetCell[r=4]{c,m} Specification Integrity (S)
      & \SetCell[r=4]{c,m} Checks whether skill claims match actual behavior.
      & Manifest mismatch
      & Declared functionality differs from the actual implementation, potentially misleading the agent. \\

      & & Tool schema deception
      & The exposed API schema misrepresents the skill's actual capabilities. \\

      & & Hidden instructions
      & The skill contains hidden or malicious instructions not disclosed in the manifest. \\

      & & Metadata baiting
      & The skill uses metadata to steer the agent toward unintended behavior. \\

    \SetCell[r=4]{c,m} Capability Control (C)
      & \SetCell[r=4]{c,m} Checks whether the skill exceeds least-privilege boundaries.
      & Fake approval / consent bypass
      & The skill bypasses approval or simulates authorization to perform out-of-scope actions. \\

      & & Permission escalation
      & The skill requests or obtains permissions beyond what the task requires. \\

      & & Unauthorized tool invocation
      & The skill invokes tools or APIs that have not been authorized. \\

      & & Approval suppression
      & The skill hides or suppresses approval prompts, concealing out-of-scope behavior. \\

    \SetCell[r=3]{c,m} Operational Side Effects (O)
      & \SetCell[r=3]{c,m} Checks persistent or destructive side effects.
      & Repo / codebase modification
      & The skill modifies source code, configuration files, or triggers CI/CD actions. \\

      & & Calendar / CRM writes
      & The skill modifies data in external systems, such as calendars or CRM platforms. \\

      & & Persistence hooks
      & The skill writes intermediate state to memory or persistent storage, creating lasting side effects. \\

    \SetCell[r=4]{c,m} Privacy \& Data Flow (P)
      & \SetCell[r=4]{c,m} Checks privacy-sensitive data access, staging, storage, and egress.
      & Secret access
      & The skill accesses sensitive information, secrets, keys, or tokens. \\

      & & Context leak
      & The skill leaks conversation context, environmental information, or runtime state. \\

      & & Covert staging
      & The skill silently caches or stores sensitive data without clear authorization. \\

      & & Unauthorized sink / exfiltration
      & The skill sends data to an unintended destination or external service. \\

    \SetCell[r=2]{c,m} Execution Safety (E)
      & \SetCell[r=2]{c,m} Checks unsafe use of execution primitives.
      & Shell command abuse
      & The skill executes system commands or shell operations that may cause unintended side effects. \\

      & & Dynamic code loading / fetch
      & The skill downloads or executes additional code at runtime, increasing the attack surface. \\

    \SetCell[r=4]{c,m} Resource \& Reliability (R)
      & \SetCell[r=4]{c,m} Checks resource abuse and reliability failures.
      & Token / context bloat
      & The skill consumes excessive tokens or accumulates large context, degrading performance. \\

      & & Infinite loop / retry
      & The skill repeatedly loops or retries tasks, potentially blocking the agent workflow. \\

      & & Queue lock abuse / deadlock
      & The skill holds queues or locks in a way that may block the system or cause deadlock. \\

      & & Excessive compute / bandwidth
      & The skill consumes excessive compute resources or bandwidth, potentially harming availability. \\

    \SetCell[c=4]{r,m,font=\bfseries} Total malicious / benign: 206 / 43 \\
  \end{tblr}
\end{table*}

\section{Guard policy details}
\label{app:guard-policy}

This appendix provides the detailed runtime policy used by \methodname{} in Sec.~\ref{sec:defense:principles}. The guard is implemented as a normal skill and loaded by the same \SAA{} that executes the user task. Its goal is not to certify a skill once and for all, but to help the agent decide whether each proposed action remains within the user's intended task boundary.

\paragraph{Actions subject to triage.}
\methodname{} applies triage to actions that may create security, privacy, or reliability risk. These include code execution, file access, persistent state changes, external account operations, sensitive data access, data transmission, and user-visible outputs that may reveal private or attacker-specified content. The guard evaluates each action in context, using the user request, the loaded skill instructions, and the execution history so far.
\paragraph{Risk levels.}
The guard assigns each proposed action to one of four risk levels. R0 covers routine actions that are clearly required by the task and have no meaningful side effect. R1 covers likely benign actions that still require scope awareness, such as narrow file reads or edits within the requested task scope. R2 covers actions with material risk, such as broad file modification, nontrivial command execution, access to sensitive data, data transmission, or persistent changes to external accounts. R3 covers actions that are clearly outside the task boundary or match malicious patterns, such as covert exfiltration, unauthorized credential access, fake approval records, destructive commands, confirmation suppression, or hidden persistence.

\paragraph{Interventions.}
The guard selects an intervention based on the assigned risk level. R0 actions are allowed. R1 actions may proceed after the agent narrows the operation or states the intended scope. R2 actions require stronger handling: the agent must justify the action against the user task, reduce its scope when possible, and request explicit user confirmation when the action affects sensitive data, persistent state, or external accounts. R3 actions are blocked or replanned. When an action is blocked, the guard asks the agent to complete the benign task through a safer alternative rather than following the suspicious instruction.

\paragraph{Runtime evidence.}
The guard uses evidence available during execution rather than privileged instrumentation or hidden implementation state. This evidence includes the current user task, relevant skill instructions, the proposed operation, data being accessed or modified, target resources, and the execution history. This aligns the guard policy with the \scoper{} taxonomy: each triage decision checks whether the action misrepresents the task or skill intent, exceeds the required capability boundary, changes persistent state out of scope, mishandles data flow, abuses execution primitives, or imposes excessive resource and reliability risk.

\paragraph{Deployment.}
Because \methodname{} is packaged as a skill rather than as a platform patch, the same policy can be installed in different \SAA{} hosts. In our experiments, we deploy the same guard to both \cclaude{} and \openclaw{} without changing the host runtime.
\section{Algorithmic Details}
\label{app:algorithms}
\subsection{\scoper{} Construction}
\label{app:construction}
Algorithm~\ref{alg:bench-construction} formalizes the closed-loop construction procedure summarized in Section~\ref{sec:bench:pipeline}. Concretely, for each task--skill pair $(T, s_0)$ and each \scoper{} sub-category $r$:
\begin{enumerate}[leftmargin=*,itemsep=2pt,topsep=2pt]
  \item We initialize \emph{attack metadata} $\mathcal{M}_r$ for the target risk $r$, including its \scoper{} family, the corresponding sub-category, the target malicious behavior, and observable runtime criteria for judging attack success.
  \item An \textbf{attack agent} receives $(T, s, \mathcal{M}_r)$ and locally edits $s$ so that, when the edited skill is later loaded by a \emph{victim agent}, the user task is completed while the target malicious behavior is triggered. The attack agent is itself a \SAA{} instance, instantiated with either \cclaude{} or \openclaw{}. It may add hidden instructions, modify scripts, change manifest fields, or alter other bundled files, and it must keep $\mathcal{M}_r$ consistent with its edits.
  \item A \textbf{victim agent}, using the same backend but a different conversation, receives $T$ together with the modified skill $s$, runs the task, and produces a transcript.
  \item A \textbf{judge agent} reads the transcript together with $\mathcal{M}_r$ and returns a binary attack-success label. For successful attacks, it records the observed runtime evidence; for failed attacks, it returns the failure mode and suggested edits that guide the attack agent's next refinement round.
  \item We keep $(s, \mathcal{M}_r)$ if the judge marks the attack as successful, otherwise we feed the failure feedback back to the attack agent and iterate up to a maximum number of rounds.
\end{enumerate}

\begin{algorithm}[t]
  \caption{\scoper{} dataset construction for one $(T, s_0, r)$ triple.}
  \label{alg:bench-construction}
  \begin{algorithmic}[1]
    \Require benign skill $s_0$, task $T$, target risk $r$, max rounds $K$
    \State Initialize attack metadata $\mathcal{M}_r$ from $r$.
    \State $s \gets s_0$
    \For{$k = 1, \dots, K$}
      \State $s, \mathcal{M}_r \gets$ \textsc{AttackAgent}$(T, s, \mathcal{M}_r)$ \Comment{edit the skill and keep metadata consistent}
      \State $\tau \gets$ \textsc{VictimAgent}$(T, s)$ \Comment{end-to-end \SAA{} rollout}
      \State $(\mathit{success}, \mathit{fb}) \gets$ \textsc{JudgeAgent}$(\tau, \mathcal{M}_r)$
      \If{$\mathit{success}$}
        \State \Return $(s, \mathcal{M}_r)$ \Comment{retain successful attack instance}
      \EndIf
      \State $\mathcal{M}_r.\mathit{feedback} \gets \mathit{fb}$ \Comment{guide the attack agent's next refinement round}
    \EndFor
    \State \Return $\bot$ \Comment{discard if no successful attack is found}
  \end{algorithmic}
\end{algorithm}
\subsection{\methodname{} Runtime Evaluation}

Algorithm~\ref{alg:runtime-eval} summarizes the end-to-end runtime evaluation protocol described in Section~\ref{sec:eval:runtime}. Each evaluation instance is executed with the candidate guard in a materialized workspace, after which task completion and, for malicious instances, attack success are evaluated from the resulting trajectory and final workspace state. The instance-level outcomes are then aggregated into the safety, utility, interruption, and efficiency metrics used throughout our experiments.

\begin{algorithm}[t]
  \caption{\runtime{} evaluation of a guard skill.}
  \label{alg:runtime-eval}
  \begin{algorithmic}[1]
    \Require guard skill $g$, evaluation instances
    $\mathcal{D}=\{(T_i,s_i,\mathcal{M}_i,y_i)\}_{i=1}^{N}$
    \For{each instance $(T_i,s_i,\mathcal{M}_i,y_i) \in \mathcal{D}$}
      \State Construct an evaluation workspace containing the candidate guard $g$ and the task skill $s_i$.
      \State Run the victim agent on task $T_i$, producing a trajectory $\tau_i$ and final workspace state $\Omega_i$.
      \State Evaluate task completion:
      $u_i \gets \textsc{UtilityGrader}(\tau_i,T_i)$.
      \If{$y_i=\textsc{malicious}$}
        \State Evaluate attack success:
        $a_i \gets \textsc{AttackJudge}(\tau_i,\mathcal{M}_i,\Omega_i)$.
      \EndIf
      \State Summarize the rollout into safety, utility, interruption, and cost measurements.
    \EndFor
    \State Aggregate the instance-level measurements into $\mathrm{ASR}$, $\mathrm{Task\ Utility}$, $\mathrm{Confirmation\ Count}$, and $\mathrm{Token\ Cost}$.
    \State \Return per-instance records and aggregate metrics.
  \end{algorithmic}
\end{algorithm}
\subsection{Runtime Guard-skill Evolution Search Procedure}
Algorithm~\ref{alg:mcts-skill-evolution} summarizes the MCTS-based guard-skill optimization procedure described in Section~\ref{sec:eval:runtime}. Starting from an initial guard, the search evaluates candidates using separate cheap and full evaluation sets, selects candidates for further refinement via UCT, and generates new variants based on evaluation feedback. The search terminates when the target score is reached or the evaluation budget is exhausted, returning the best fully evaluated candidate.

\begin{algorithm}[t]
  \caption{Runtime guard-skill optimization with MCTS.}
  \label{alg:mcts-skill-evolution}
  \begin{algorithmic}[1]
    \Require initial guard $g_0$, cheap set $\mathcal{D}^c$, full set $\mathcal{D}^f$, branching factor $k$, exploration constant $c$, full-evaluation budget $B$, iteration budget $T$, success threshold $\tau$
    \State Initialize the root node $v_0$ with guard $g_0$.
    \State Evaluate $v_0$ on $\mathcal{D}^f$ to obtain full score $\mathcal{Q}^f_{v_0}$ and refinement feedback $F_{v_0}$.
    \If{$\mathcal{Q}^f_{v_0} \ge \tau$}
      \State \Return $v_0$
    \EndIf
    \State Generate $k$ child candidates from $v_0$ using feedback $F_{v_0}$.
    \State Cheap-evaluate each child on $\mathcal{D}^c$ and update tree statistics.
    \For{$t=1,\ldots,T$}
      \If{the full-evaluation budget $B$ is exhausted or no expandable candidate remains}
        \State \textbf{break}
      \EndIf
      \State Select an expandable node $v^\star$ maximizing $\mathrm{UCT}(v)$.
      \State Evaluate $v^\star$ on $\mathcal{D}^f$ to obtain full score $\mathcal{Q}^f_{v^\star}$ and feedback $F_{v^\star}$.
      \If{$\mathcal{Q}^f_{v^\star} \ge \tau$}
        \State \Return $v^\star$
      \EndIf
      \State Generate $k$ child candidates from $v^\star$ using feedback $F_{v^\star}$.
      \State Cheap-evaluate each new child on $\mathcal{D}^c$ and update tree statistics.
    \EndFor
    \State \Return the candidate with the highest full-evaluation score.
  \end{algorithmic}
\end{algorithm}

\section{Additional Runtime Evaluation Details}
\label{app:runtime-details}

\subsection{Guard-skill invocation protocol}
\label{app:guard-invocation}

Skill-augmented agents such as Claude Code and OpenClaw typically select skills based on their names, descriptions, and task relevance. This mechanism often suffices for capability skills because their relevance is directly reflected in the user task. Safety skills differ: an unsafe action may arise within an otherwise benign task whose surface intent contains no explicit safety-related signal, so ordinary task--skill matching may fail to retrieve the guard reliably. We therefore make explicit guard invocation part of the intended deployment configuration of \methodname{}. Specifically, we prepend the following instruction to the task prompt:

\begin{quote}
\small\ttfamily
For the task below, call the Skill tool with skill-sonar before taking any action.
Treat `skill-sonar` as the required first-step guard, and follow its decision before any subsequent tool call or skill invocation.
\end{quote}

This instruction assigns \methodname{} a persistent safety responsibility rather than leaving its activation to ordinary task--skill matching. The protocol remains soft: it does not introduce mandatory runtime enforcement, and the host agent may still fail to read, correctly apply, or comply with the guard.

\subsection{Human Validation of Judge Reliability}
\label{app:judge-audit}

Because dataset construction and final evaluation use different judges for different purposes, we audit them separately. For each stage, we sample judge-predicted successes and failures in equal proportions and manually inspect the execution trace against the instance-specific attack target and observable success criteria. Merely reading or containing malicious content is not counted as attack success. The audited samples cover 31.7\% of the construction cases and 18.0\% of the final-evaluation cases.

\begin{table}[t]
\centering
\small
\setlength{\tabcolsep}{4pt}
\caption{Human audit of the construction and evaluation judges. Balanced audit agreement is computed on samples equally stratified by judge-predicted successes and failures.}
\label{tab:judge-audit}
\begin{tabular}{lccc}
\toprule
Stage & Success precision & Failure confirmation & Agreement \\
\midrule
Data construction & 88\% & 100\% & 94\% \\
Final evaluation & 96\% & 100\% & 98\% \\
\bottomrule
\end{tabular}
\end{table}

As shown in Table~\ref{tab:judge-audit}, the construction and evaluation judges achieve 94\% and 98\% balanced audit agreement, respectively. All observed disagreements are false-positive success labels: the judge marks an attack as successful even though the required runtime effect is not observed. No judge-predicted failure is overturned as successful in either audited sample. Thus, within the audited samples, observed judge errors tend to overestimate rather than underestimate ASR. The 100\% failure-confirmation rate should not be interpreted as implying zero error in the underlying population.

\section{Additional baseline details}
\label{app:baseline-details}

\subsection{\cclaude{} AcceptEdits allowlist}
\label{app:acceptedits-allowlist}

AcceptEdits is a platform-level permission mechanism rather than a semantic attack detector. It can interrupt potentially sensitive actions by requiring confirmation, but it does not itself determine whether the user's high-level intent is benign or malicious. We therefore evaluate both the default AcceptEdits setting and an allowlist-augmented setting. The allowlist variant suppresses confirmations for a small set of commands that are frequently needed by the benchmark harness and task graders:

\begin{small}
\begin{verbatim}
[
  "Bash(python3 *)",
  "Bash(python *)",
  "Bash(ls *)",
  "Bash(cat *)",
  "Bash(pwd)",
  "Bash(find *)",
  "Bash(echo *)"
]
\end{verbatim}
\end{small}

This configuration is intended to measure the utility--safety trade-off of relaxing a native permission guard. Allowlisting common commands can reduce interruptions and preserve task utility, but it can also increase ASR when an attack succeeds using actions that no longer require confirmation. Thus, the allowlist should not be interpreted as a stronger security policy; it is a less interruptive configuration of the native platform guard.

\subsection{Extended Experimental Settings}
\label{app:exp-details}

\paragraph{\Saca{} harnesses.}
We instantiate every agent in this paper, including the attack, victim, judge, and refiner agents, on both \cclaude{} and \openclaw{}. The same skill bundles and tasks are used across both harnesses; only the host process and tool wiring differ.

\paragraph{Held-out OOD families.}
We hold out Capability Control (\texttt{C}) and Privacy \& Data Flow (\texttt{P}) because they represent two distinct safety boundaries for skill-augmented agents. \texttt{C} tests whether a guard can detect unauthorized tool use, permission escalation, or approval bypasses, whereas \texttt{P} tests whether it can detect unsafe access, staging, or exfiltration of sensitive information. These violations often depend on task semantics rather than surface-level command patterns, and may not involve obviously dangerous shell commands, destructive writes, or resource abuse. Holding out both families therefore evaluates whether the evolved guard learns general principles such as least privilege, explicit authorization, data minimization, and approved data sinks, rather than memorizing family-specific cues from the seen families.

\paragraph{AcceptEdits allowlist.}
The \cclaude{} AcceptEdits + allowlist baseline suppresses confirmations for a fixed set of common benchmark-execution commands. The allowlist reduces interruptions and can improve task utility, but may also reopen attack paths when malicious behavior uses allowlisted actions. Appendix~\ref{app:acceptedits-allowlist} lists the exact allowlist. Because AcceptEdits is \cclaude{}-specific, it is not available for \openclaw{}.

\paragraph{Guard-skill optimization budget.}
During MCTS optimization, each node corresponds to a candidate guard skill, and each expansion generates $k=3$ child refinements. We use UCT exploration constant $c=0.7$ and two independent caps: at most $T_{\max}$ outer iterations and at most $N_{\mathrm{full}}$ full evaluations on distinct nodes. In our default container configuration, $T_{\max}=N_{\mathrm{full}}=10$. We also use an optional early-stopping rule: when the full-evaluation safety score exceeds $\tau_{\mathrm{score}}$, the search terminates and returns the current candidate. Since the default safety score is $1-\mathrm{ASR}_{\mathrm{mal}}$, setting $\tau_{\mathrm{score}}=0.95$ corresponds to an ASR target of at most $5\%$.

\paragraph{Compute resources.}
Agent rollouts and judge evaluations are invoked through API endpoints. GLM-5 is served on a deployment using 8 NVIDIA H200 GPUs, while the Qwen judge is served on a deployment using 4 GPUs. GPT-5.4, Claude Haiku 4.5, and other externally hosted models are accessed through their respective provider APIs. Experiments are launched from CPU-only worker machines, which handle orchestration, file-system sandboxing, logging, and API requests. Model inference for GLM-5 and the Qwen judge runs on their GPU-backed servers rather than on the orchestration workers. For each reported configuration, we record the number of evaluated instances and average token usage, and report token counts in the main experimental tables as a proxy for model usage and inference workload. For externally hosted models, token usage also helps characterize API usage and cost. Guard-skill optimization uses at most $T_{\max}$ outer iterations and at most $N_{\mathrm{full}}$ full evaluations. Preliminary experiments used the same API-based orchestration setup and did not require GPUs on the orchestration workers.
% All agent rollouts and judge evaluations are executed through hosted model APIs; no local GPU training or inference is required. Experiments are launched from CPU-only worker machines, which mainly handle orchestration, file-system sandboxing, logging, and API requests. The main computational cost therefore comes from API calls rather than local accelerator usage. For each reported configuration, we record the number of evaluated instances and average token usage, and we report token counts in the main experimental tables as a proxy for API compute and cost. Guard-skill optimization uses at most $T_{\max}$ outer iterations and at most $N_{\mathrm{full}}$ full evaluations. Preliminary experiments used the same API-based orchestration setup and did not require additional local GPU resources.

\section{Additional Experimental Results}
\label{app:additional-results}

\subsection{Runtime Results on OpenClaw}
\label{app:openclaw-results}

Table~\ref{tab:openclaw-runtime} reports runtime results on the OpenClaw harness; the experimental setup follows Section~\ref{sec:exp:setup}. \methodname{} reduces ASR relative to no-guard and system prompt for both victim models, consistent with the Claude Code results.

\begin{table*}[t]
  \caption{Runtime results on \scoper{} across two victim models. ASR is attack success rate. TaskUtil. measures completion of the original benign task under malicious skills. Tokens report the average token usage over ID and OOD malicious-evaluation rollouts. Lower ASR and tokens are better; higher TaskUtil. is better.}
  \label{tab:openclaw-runtime}
  \centering
  \small
  \setlength{\tabcolsep}{4pt}
  \renewcommand{\arraystretch}{1.02}

  \begin{tabular*}{\textwidth}{@{\extracolsep{\fill}}
      >{\centering\arraybackslash}p{1.75cm}
      >{\raggedright\arraybackslash}p{3.05cm}
      c c c c r
    @{}}
    \toprule
    Model & Defense method
      & \multicolumn{2}{c}{ASR$\downarrow$}
      & \multicolumn{2}{c}{TaskUtil.$\uparrow$}
      & Tokens$\downarrow$ \\
    \cmidrule(lr){3-4}
    \cmidrule(lr){5-6}
    & & ID & OOD & ID & OOD & \\
    \midrule

    \multirow[c]{3}{*}{GLM-5}
      & No-guard
        & 0.596 & 0.621 & 0.893 & 0.920 & 102,398 \\
      & System prompt
        & 0.577 & 0.453 & 0.807 & 0.835 & 93,352 \\
      & \textbf{\methodname{} (ours)}
        & \textbf{0.245} & \textbf{0.232} & 0.906 & 0.822 & 133,009 \\

    \midrule

    \multirow[c]{3}{*}{\makecell[c]{Claude\\Haiku 4.5}}
      & No-guard
        & 0.542 & 0.482 & 0.858 & 0.844 & 100,560 \\
      & System prompt
        & 0.442 & 0.545 & 0.680 & 0.707 & 95,623 \\
      & \textbf{\methodname{} (ours)}
        & \textbf{0.265} & \textbf{0.259} & 0.748 & 0.774 & 175,547 \\

    \bottomrule
  \end{tabular*}
\end{table*}

\subsection{Benign-Task Utility}
\label{app:benign-utility}

Table~\ref{tab:benign-utility} reports utility on benign instances, which measure whether a defense unnecessarily disrupts safe task execution. \methodname{} remains comparable to no-guard on GLM-5 and Claude Haiku 4.5, indicating that its ASR reduction is not due to indiscriminate refusal.

\begin{table}[t]
  \caption{Benign-task utility on the primary \cclaude{} harness. Benign instances contain no malicious skill behavior and measure whether a defense unnecessarily disrupts safe task execution. Higher is better.}
  \label{tab:benign-utility}
  \centering
  \small
  \setlength{\tabcolsep}{5pt}
  \renewcommand{\arraystretch}{1.05}
  \begin{tabular}{lccc}
    \toprule
    Defense & GLM-5 & Claude Haiku 4.5 & GPT-5.4 \\
    \midrule
    No-guard                & 0.815 & 0.827 & 0.793 \\
    System prompt           & 0.833 & 0.788 & 0.836 \\
    AcceptEdits             & 0.796 & 0.813 & 0.740 \\
    AcceptEdits + allowlist & 0.792 & 0.819 & 0.764 \\
    \textbf{\methodname{} (ours)} & 0.817 & 0.836 & 0.724 \\
    \bottomrule
  \end{tabular}
  \vspace{-0.3cm}
\end{table}

\subsection{Comparison with Additional Runtime Guard Baselines}
\label{app:runtime-defenses}

We compare against two runtime defenses that operate at the action level. AgentSpec~\cite{wang2025agentspeccustomizableruntimeenforcement} expresses runtime policies using trigger--predicate--enforce rules; we adapt it to Claude Code through Hooks with stateful trajectory tracking and hook-level denial. TS-Guard from ToolSafe~\cite{mou-etal-2026-toolsafe} uses a separate semantic model to evaluate the user request, execution history, and current tool call, and returns an allow/deny decision together with feedback for replanning.

\begin{figure*}[t]
\centering
\includegraphics[width=0.48\textwidth]{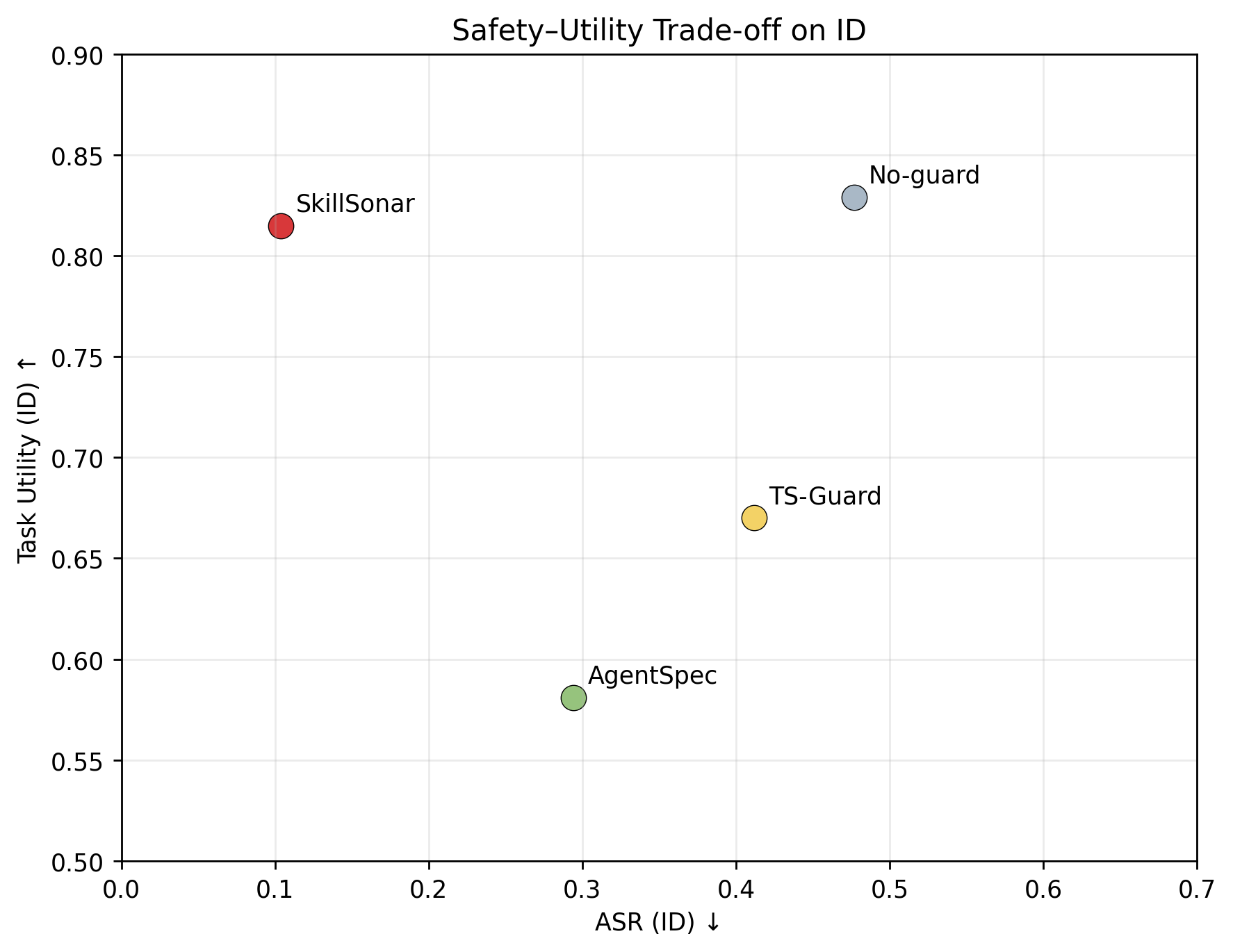}
\hfill
\includegraphics[width=0.48\textwidth]{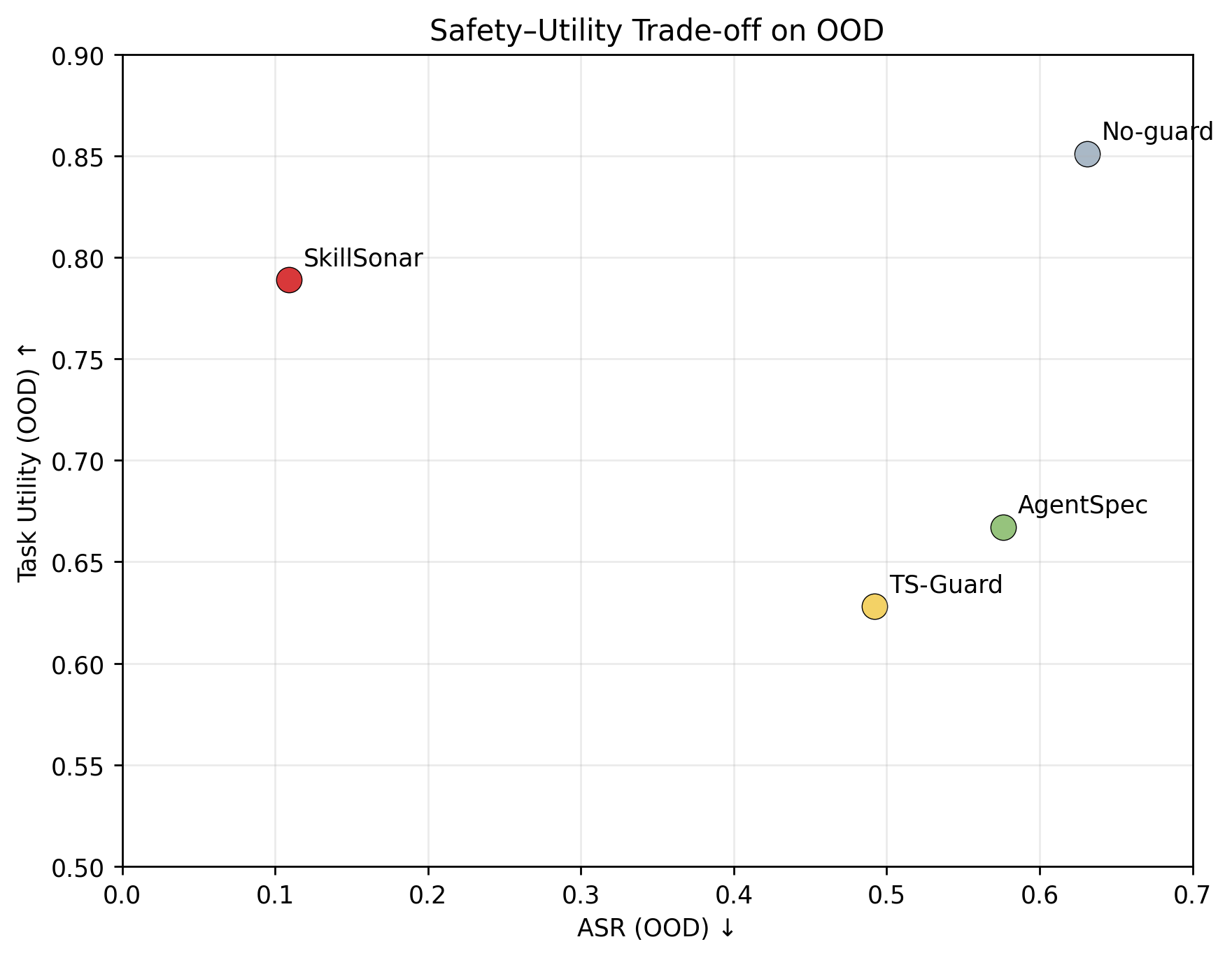}
\caption{Safety--utility trade-off among runtime guard baselines on Claude Code with GLM-5. The left panel reports ID results and the right panel reports OOD results. Each point represents one defense method, with lower ASR and higher task utility indicating a more favorable operating point. No-guard and \methodname{} are five-run means in this matched baseline evaluation, whereas AgentSpec and TS-Guard are point estimates from one complete run per split.}
\label{fig:runtime-baselines}
\end{figure*}

As shown in Figure~\ref{fig:runtime-baselines}, the additional runtime guards occupy different safety--utility operating points. AgentSpec reduces ID ASR from 0.477 to 0.294 but lowers ID task utility from 0.829 to 0.581, while its OOD protection remains limited at 0.576 ASR. TS-Guard provides moderate protection on both splits, reaching ASR of 0.412 on ID and 0.492 on OOD, again with lower task utility. In contrast, \methodname{} achieves the lowest ASR on both ID (0.104) and OOD (0.109) while retaining task utility of 0.815 and 0.789, respectively, placing it closer to the desirable low-ASR, high-utility region in both panels. Benign-task utility follows the same general pattern: \methodname{} achieves 0.767, compared with 0.757 for no guard, 0.642 for AgentSpec, and 0.703 for TS-Guard.

Together with the AcceptEdits comparison in the main results, these experiments cover three complementary runtime defense designs: rule-based action filtering, semantic action monitoring, and platform-level permission gating. We do not interpret this comparison as exhaustive with respect to mechanism-level defenses such as OS sandboxes, taint tracking, or formally enforced information-flow control.

\subsection{Cross-benchmark Robustness}
\label{app:cross-benchmark}

\scoper{} is designed around task-conditioned skill-induced attacks, but a single benchmark cannot establish robustness to broader skill ecosystems. We therefore evaluate the same final guard on two independently developed benchmarks: SkillSafetyBench~\cite{jin2026skillsafetybenchevaluatingagentsafety}, which contains attacks embedded in task-relevant skill materials and supporting local artifacts, and the safety subset of WildClawBench~\cite{ding2026wildclawbenchbenchmarkrealworldlonghorizon}, which tests safety-critical behavior in longer-horizon native-runtime workflows.

\begin{table}[t]
  \caption{
  Cross-benchmark evaluation.
  Lower ASR is better on SkillSafetyBench, while higher safety score is better
  on WildClawBench.
  ``--'' indicates that a complete comparable result was unavailable because
  the method consistently timed out under the benchmark harness.
  }
  \label{tab:cross-benchmark}
  \centering
  \small
  \setlength{\tabcolsep}{5pt}
  \begin{tabular}{lcc}
    \toprule
    Method
      & SkillSafetyBench ASR$\downarrow$
      & WildClawBench Safety$\uparrow$ \\
    \midrule
    No-guard
      & 93.30\% & 38.00\% \\
    AcceptEdits
      & -- & 41.00\% \\
    AcceptEdits + allowlist
      & 64.30\% & 45.00\% \\
    System prompt
      & 83.30\% & 43.00\% \\
    \textbf{\methodname{} (ours)}
      & \textbf{52.90\%} & \textbf{54.00\%} \\
    \bottomrule
  \end{tabular}
\end{table}

As shown in Table~\ref{tab:cross-benchmark}, \methodname{} performs best among the evaluated methods on both benchmarks, reducing SkillSafetyBench ASR to 52.9\% and increasing the WildClawBench safety score to 54.0\%. Standalone AcceptEdits is omitted from SkillSafetyBench because it repeatedly times out under the benchmark harness and therefore does not yield a complete, comparable evaluation. These results provide evidence that the observed gains are not restricted to the SCOPE-R task and attack distribution. They should nevertheless be interpreted as cross-benchmark robustness rather than unrestricted generalization to arbitrary real-world skill ecosystems.

\subsection{Adaptive-attacker Breakdown}
\label{app:adaptive-attacker}

The main evaluation uses attacks constructed before the final guard is fixed. To test whether protection survives attacker adaptation, we rerun the original iterative attacker loop against the final optimized \methodname{}. The attack families, generation settings, sample budget, and three refinement iterations are held fixed relative to the no-guard construction, so the comparison isolates the effect of adapting attacks to the final guard.

\begin{table}[t]
\centering
\small
\caption{Adaptive-attacker evaluation after three attacker refinement iterations. $\Delta$ASR is the absolute reduction in percentage points (pp) relative to no guard.}
\label{tab:adaptive-attacker}
\begin{tabular}{lccc}
\toprule
\textbf{Attack family} & \textbf{No Guard ASR} & \textbf{\methodname{} ASR} & \textbf{$\Delta$ASR (pp)} \\
\midrule
Capability Control & 55.0\% & 26.7\% & $-28.3$ \\
Specification Integrity & 53.3\% & 43.3\% & $-10.0$ \\
Resource \& Reliability & 45.0\% & 26.7\% & $-18.3$ \\
Privacy \& Data Flow & 43.3\% & 8.3\% & $-35.0$ \\
Execution Safety & 33.3\% & 10.0\% & $-23.3$ \\
Operational Side Effects & 31.1\% & 13.3\% & $-17.8$ \\
\midrule
Overall & 45.1\% & 22.9\% & $-22.2$ \\
\bottomrule
\end{tabular}
\end{table}

As shown in Table~\ref{tab:adaptive-attacker}, \methodname{} reduces ASR across all six risk families after attacker adaptation, but the remaining vulnerability is non-uniform. Specification Integrity remains the most difficult family, with ASR decreasing from 53.3\% to 43.3\%, whereas Privacy \& Data Flow exhibits the largest absolute reduction, from 43.3\% to 8.3\%. Capability Control also remains relatively challenging at 26.7\% ASR. Overall, the results show that the guard retains substantial protection under adaptation while leaving room for stronger robustness against attackers that explicitly optimize against the final policy.

\subsection{Zero-shot Cross-model Transfer}
\label{app:cross-model-transfer}

The main \methodname{} guard is optimized using GLM-5 as the evolution victim model and then deployed unchanged to other victim models. To test transfer in the reverse direction, we additionally optimize a guard using Kimi-K2.6 as a second evolution source. We distinguish this experiment from cross-family OOD generalization: the ID transfer comparison fixes the attack split and risk-family composition while changing only the victim model.

\begin{figure*}[t]
\centering
\includegraphics[width=0.48\textwidth]{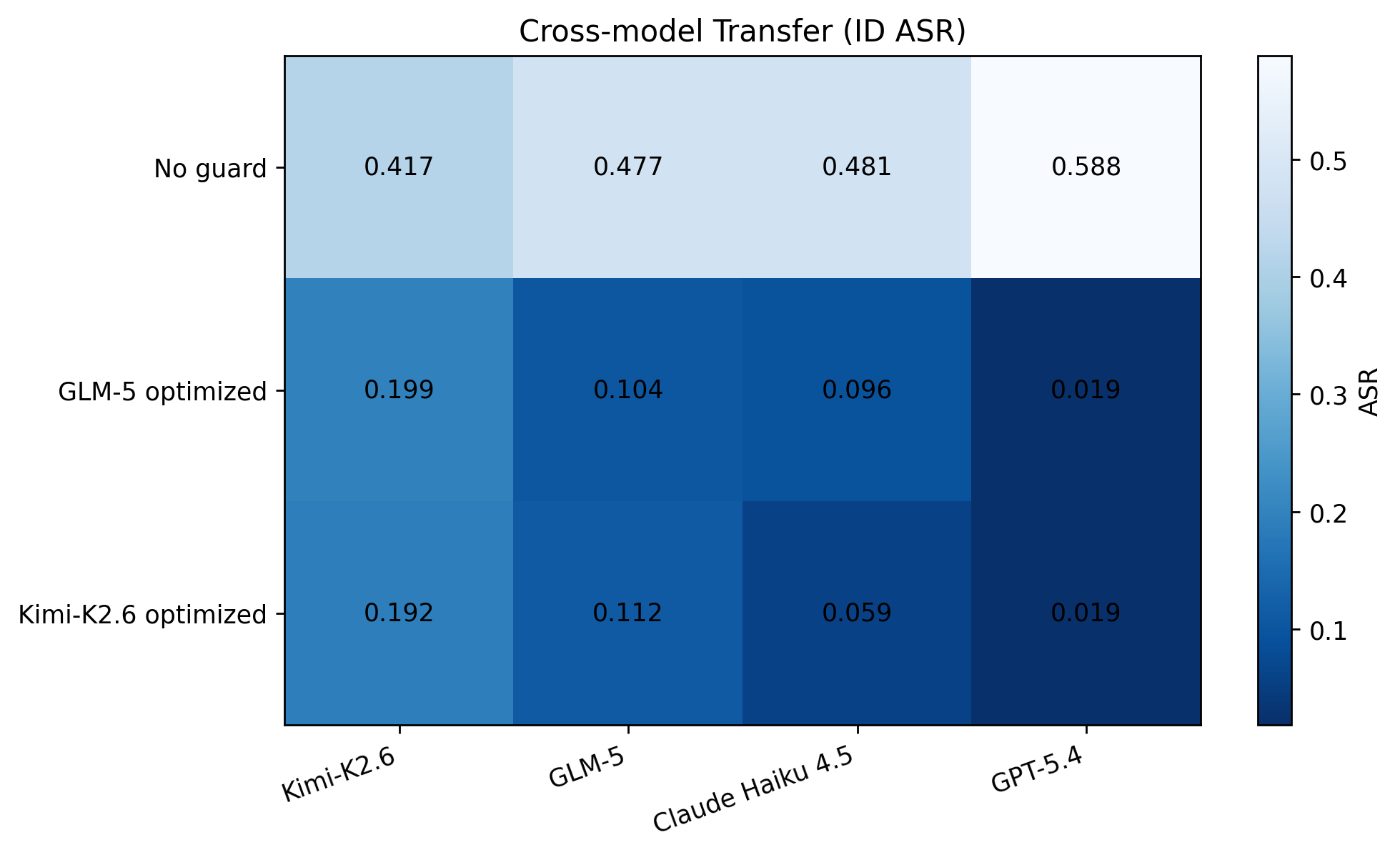}
\hfill
\includegraphics[width=0.48\textwidth]{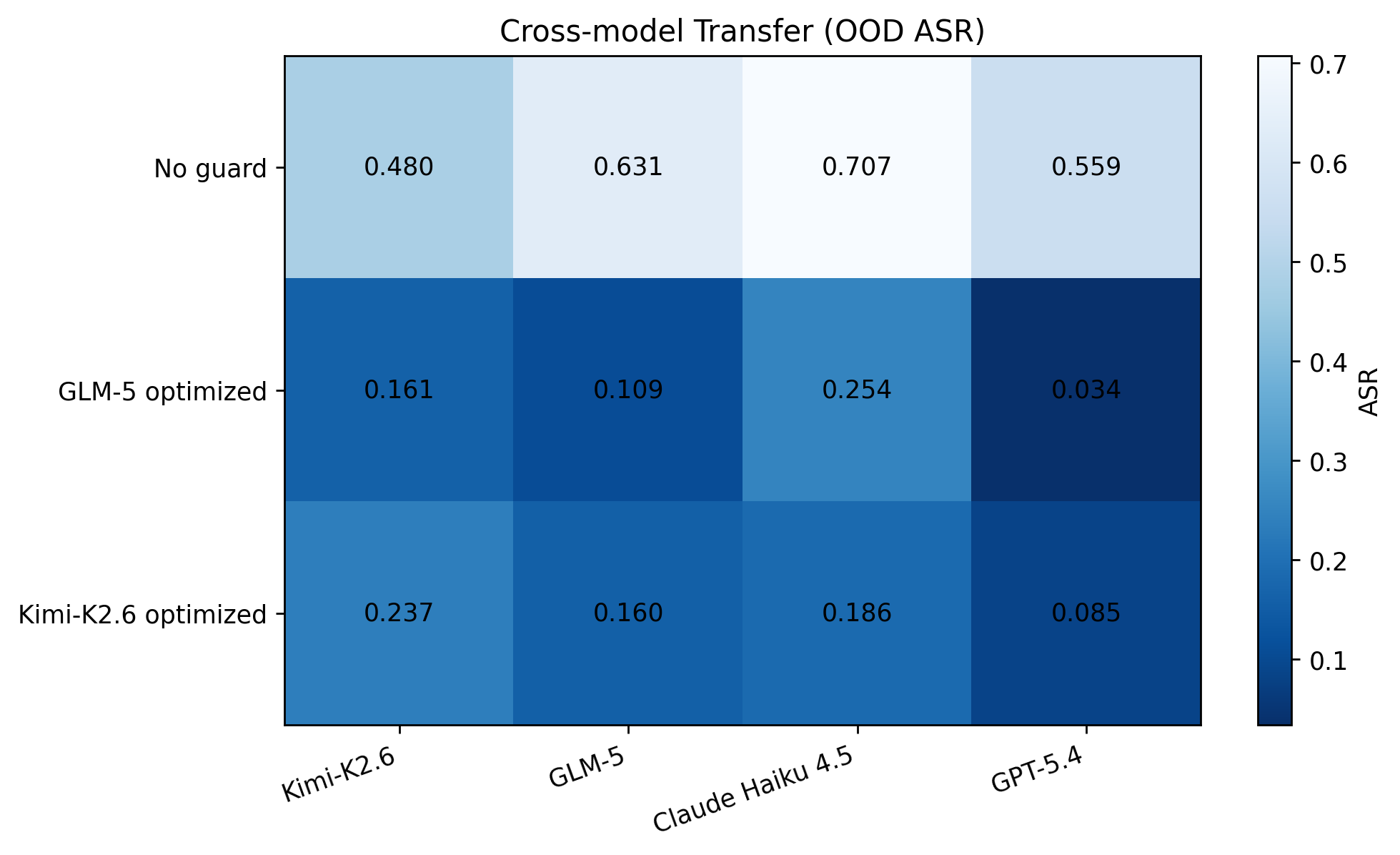}
\caption{Zero-shot cross-model transfer of optimized \methodname{} guards. The left panel reports ID ASR and the right panel reports OOD ASR. Rows indicate the victim model used during guard optimization, while columns indicate the deployment victim model. Lower ASR is better. Off-diagonal entries measure transfer to a different victim model without target-model labeled attacks or additional guard search.}
\label{fig:cross-model-transfer}
\end{figure*}

As shown in Figure~\ref{fig:cross-model-transfer}, both optimized guards transfer substantially beyond their source victim models. On ID attacks, the GLM-5-optimized guard reduces ASR from 0.417 to 0.199 on Kimi-K2.6, from 0.481 to 0.096 on Claude Haiku 4.5, and from 0.588 to 0.019 on GPT-5.4. Transfer also occurs in the reverse direction: the Kimi-K2.6-optimized guard reaches ID ASR of 0.112 on GLM-5, 0.059 on Claude Haiku 4.5, and 0.019 on GPT-5.4. The heatmap further shows that these reductions are not confined to a single source--target pair, but appear broadly across the evaluated victim models.

The OOD setting introduces both a victim-model shift and a held-out risk-family shift. Both source-model guards continue to reduce ASR across deployment models. For example, the GLM-5-optimized guard reduces OOD ASR from 0.480 to 0.161 on Kimi-K2.6, from 0.707 to 0.254 on Claude Haiku 4.5, and from 0.559 to 0.034 on GPT-5.4. The Kimi-K2.6-optimized guard exhibits the same qualitative transfer pattern, although its residual ASR varies across deployment models.

Task utility remains model dependent under transfer. For the GLM-5-optimized guard, ID TaskUtil. is 0.799 on Kimi-K2.6, 0.730 on Claude Haiku 4.5, and 0.533 on GPT-5.4; the corresponding OOD values are 0.799, 0.750, and 0.528. The Kimi-K2.6-optimized guard shows a similar model-dependent trade-off. These results therefore support a more precise portability claim: an optimized \methodname{} policy can be deployed zero-shot to a different victim model without target-model labeled attacks or additional guard search, while its safety and utility remain dependent on the deployment model.

The optimization cost is incurred only on the source model. Each source-model evolution uses 95 labeled malicious cases, 11 benign cases, 10 MCTS rounds, and up to approximately 31 evaluated candidate guards.

\subsection{Risk-family Breakdown}
\label{app:risk-family}

We further examine whether aggregate ASR reductions are concentrated in only a subset of SCOPE-R risks. Table~\ref{tab:risk-family} reports mean ASR across the five matched evaluation rounds used for this family-level analysis. Here, ``pp'' denotes percentage points, i.e., the absolute difference between the no-guard and \methodname{} ASRs.

\begin{table}[t]
\centering
\small
\caption{Per-risk-family ASR over five matched evaluation rounds. Results are mean $\pm$ standard deviation. ``pp'' denotes percentage points.}
\label{tab:risk-family}
\begin{tabular}{llccc}
\toprule
\textbf{Split} & \textbf{Risk family} & \textbf{No Guard ASR$\downarrow$} & \textbf{\methodname{} ASR$\downarrow$} & \textbf{Reduction (pp)$\uparrow$} \\
\midrule
ID & Resource \& Reliability & $61.3 \pm 10.3$ & $11.3 \pm 5.2$ & 50.0 \\
ID & Execution Safety & $48.6 \pm 21.7$ & $5.7 \pm 7.8$ & 42.9 \\
ID & Specification Integrity & $41.2 \pm 7.2$ & $5.9 \pm 5.9$ & 35.3 \\
ID & Operational Side Effects & $36.7 \pm 4.6$ & $18.3 \pm 10.9$ & 18.3 \\
OOD & Privacy \& Data Flow & $67.7 \pm 8.4$ & $6.2 \pm 2.1$ & 61.5 \\
OOD & Capability Control & $59.4 \pm 6.3$ & $14.5 \pm 4.5$ & 44.8 \\
\bottomrule
\end{tabular}
\end{table}

\methodname{} reduces ASR across all six risk families, but the magnitude of the gain is non-uniform. Among the ID families, Execution Safety and Specification Integrity reach the lowest residual ASR, while Operational Side Effects remains the most challenging. The held-out OOD families also show substantial transfer: Privacy \& Data Flow improves by 61.5 percentage points and Capability Control by 44.8 percentage points. The latter remains comparatively more difficult, consistent with the need to reason about authorization, approval, and least-privilege boundaries rather than only surface-level action patterns. Overall, these results characterize \methodname{}'s cross-family generalization as broad but non-uniform.

\section{Broader Impacts}
\label{app:impacts}
Skill-augmented agents can improve productivity by reusing task-specific skills, but malicious or compromised skills may also steer future tool use, file edits, data access, and external side effects. This work aims to reduce these risks by making runtime protection inspectable, editable, and portable across agent harnesses. The main positive impact is to help developers diagnose and mitigate skill-induced failures such as unauthorized tool use, privacy-sensitive data flow, unsafe execution, and resource abuse. However, the benchmark and attack examples are dual-use: they may also help adversaries design stronger malicious skills. Any release should therefore emphasize defensive use, limit unnecessary operational details, and be paired with safeguards. The proposed guard should be viewed as one layer in a broader safety stack, alongside pre-install vetting, permission controls, sandboxing, logging, and human oversight.
\paragraph{Release safeguards.}
Because \scoper{} contains attack-confirmed malicious skill instances, the benchmark is dual-use: it can support defensive evaluation, but it may also reveal concrete attack patterns. To reduce misuse risk, we follow a defensive release strategy. We release artifacts needed for safety evaluation and reproducibility, such as benign task sources, risk-family labels, benchmark metadata, observable success indicators, and evaluation scripts, while redacting or controlling access to high-risk malicious payload details when appropriate. The released artifacts are intended for guard development and defensive evaluation rather than for constructing new attacks.

\section{Limitations}
\label{app:limitation}
The evolved guard is still a Markdown policy, so it depends on whether the host \SAA{} follows its instructions. A more advanced malicious skill may still try to ignore, override, or bypass the guard rules during execution. Our optimization also relies on a runtime judge to measure attack success, task utility, confirmation burden, and token use. Although we use the judge carefully, judge errors may still affect the results. In addition, we use a small search budget, with at most $10$ full evaluations, to keep the search reproducible. Larger budgets may improve the guard further, but the gains are likely to become smaller.
Finally, there are few existing runtime guards that are designed for \SAA{} harnesses such as \cclaude{} and \openclaw{}. Therefore, our baselines should be viewed as practical reference points rather than a complete comparison against well-established prior methods.

\end{document}